\documentclass[reprint,amsmath,amssymb,aps, prd]{revtex4-2}
\usepackage[colorlinks=true,linkcolor=blue,citecolor=blue,urlcolor=blue]{hyperref}
\usepackage{newtxtext,newtxmath}
\usepackage[T1]{fontenc}
\usepackage{ae,aecompl}
\usepackage[dvipsnames]{xcolor}
\usepackage{caption}
\usepackage{subcaption}
\usepackage{graphicx}	% Including figure files
\usepackage{amsmath}	% Advanced maths commands
\newcommand{\jcap}{J. Cosmol. Astropart. Phys.}
\newcommand{\mnras}{Mon.\ Not.\ R.\ Astron.\ Soc.}
\newcommand{\apjl}{Astrophys.\ J.\ Letters}

\newcommand{\aap}{Astron. Astrophys.}
\newcommand{\aj}{Astron. J.}

\newcommand{\physrep}{Phys. Rep.}
\newcommand{\apss}{Astrophys. Space Sci.}

\def\ln{{\rm ln}}

\def\bfx{{\bf x}}
\def\bfk{{\bf k}}
\def\bfq{{\bf q}}

\def\bfu{{\bf u}}

\def\calI{{\mathcal I}}

\def\calP{{\mathcal P}}

\def\calT{{\mathcal T}}
\def\calO{{\mathcal O}}

\renewcommand{\vec}[1]{\boldsymbol{#1}}

\def\bfk{{\bf k}}

\def\hk{{\hat{k}}}

\def\bfr{{\bf r}}
\def\hr{{\hat{r}}}

\def\calH{{\cal H}}

\def\calN{{\cal N}}
\def\calO{{\cal O}}

\def\calP{{\cal P}}
\def\calR{{\cal R}}

\def\bfx{{\bf x}}

\newcommand{\hu}{\hat u}

\newcommand{\FT}[1]{\mathcal{FT}\Big[{#1}\Big]}
\newcommand{\iFT}[1]{\mathcal{FT}^{-1}\Big[{#1}\Big]}
\newcommand{\av}[1]{\left\langle{#1}\right\rangle} 

\def\gpch{{h^{-1}\rm{Gpc}}}
\def\mpch{{h^{-1}\rm{Mpc}}}
\def\hmpc{{{\rm Mpc}^{-1} h}}

\newcommand{\six}[6]{\left(\begin{array}{ccc}
									{#1}& {#2}& {#3}\\
									{#4}& {#5}& {#6} \\
\end{array}\right)}

\begin{document}
\label{firstpage}

\newcommand{\dsj}[1]{\textcolor{Cerulean}{DSJ: #1}}

\title{Can Baryon Acoustic Oscillations Illuminate the Parity-Violating Galaxy 4PCF?} 
\author{Jiamin Hou}
\altaffiliation{jiamin.hou@mpe.mpg.de}
\affiliation{Max-Planck-Institut f{\"u}r Extraterrestrische Physik, Postfach 1312, Giessenbachstrasse, 85748 Garching, Germany}
\affiliation{Department of Astronomy, University of Florida, Gainesville, FL 32611, USA}

\author{Zachary Slepian}
\altaffiliation{zslepian@ufl.edu}
\affiliation{Department of Astronomy, University of Florida, Gainesville, FL 32611, USA}
\affiliation{Lawrence Berkeley National Laboratory, 1 Cyclotron Road, Berkeley CA 94720, USA}

\author{Drew Jamieson}
\altaffiliation{jamieson@mpa-garching.mpg.de}
\affiliation{Max-Planck-Institut f{\"u}r Astrophysik, Karl-Schwarzschild-Straße 1, 85748 Garching, Germany}

% Abstract of the paper
\begin{abstract}
Measurements of the galaxy 4-Point Correlation Function (4PCF) from theSloan Digital Sky Survey Baryon Oscillation Spectroscopic Survey (SDSS BOSS) have recently found strong statistical evidence for parity violation. If this signal is of genuine physical origin, it must stem from beyond-Standard Model physics, most likely during the very early Universe, prior to decoupling  ($z$$\sim$$1,020$). Since the Baryon Acoustic Oscillation (BAO) features imprint at decoupling, they are expected in the parity-odd galaxy 4PCF, and so detecting them would be an additional piece of evidence that the signal is genuine. We demonstrate in a toy parity-violating model how the BAO imprint on the parity-odd 4PCF. We then outline how to perform a model-independent search for BAO in the odd 4PCF, desirable since, if the signal is real, we may not know for some time what model of \textit{e.g.} inflation is producing it. If BAO are detected in the parity-odd sector, they can be used as a standard ruler as is already done in the 2PCF and 3PCF. We derive a simple formula relating the expected precision on the BAO scale  to  the overall parity-odd detection significance. Pursuing BAO in the odd 4PCF of future redshift surveys such as DESI, Euclid, Spherex, and Roman will be a valuable additional avenue to determine if parity violation in the distribution of galaxies is of genuine cosmological origin.
\end{abstract}


\maketitle
%%%%%%%%%%%%%%%%%%%%%%%%

\section{Introduction}
\label{sec:intro}

In developing fundamental theories, symmetries such as translation and rotation invariance are often used as starting points. Until the 1950s, invariance under spatial inversion (parity), was also believed to be a symmetry of nature. However,~\citep{Wu} found that this is violated in the weak interaction. Shortly thereafter, \citep{Sakharov1967} pointed out that CP (C for charge) violation is necessary to explain the observed excess of baryons over anti-baryons in the Universe. 

On cosmological scales, unlike translation and rotation symmetry, parity symmetry has not been systematically tested with observational data.  Several theoretical proposals~\citep{Carroll1990:parity, Lue1999:parity, Gluscevic2010:PVcmb, Kamionkowski2011:ParityCMBbk,Alexander2008:ParityCMB, Bartolo2015:PVcmb, Seto2007:PVGWinterferometer, Yunes2010:GWpv, Masui2017:parityGW, Jenks2023:parametrizeParity, Biagetti2020:pgwIA, Yu2020:GalaxySpinPV, Shim2024:VectorChiral,Zhu2024:parity} and observational studies~\citep{Minami2020:Birefringence,Komatsu:2022nvu,Eskilt:2023nxm, Martinovic2021:GWpv,Ng2023:AmplBirefrigenceGWTC3} have focused on two- and three-point correlations of vector and tensor quantities. Scalar quantities such as the galaxy density fluctuation field have been less associated with parity violation, as scalars are parity-conserving by nature. Indeed, to probe parity violation using scalar quantities requires at minimum a 4-Point Correlation Function (4PCF).

In the context of galaxy surveys,~\cite{Cahn2021:parity} proposed using the galaxy 4PCF to test parity violation (\citep{Shiraishi2016:PVcmb} suggested this prospect for the Cosmic Microwave Background (CMB)). The approach of \cite{Cahn2021:parity} exploits the isotropic basis functions developed in~\citep{Cahn202010} and efficient  NPCF measurement algorithms built on them~\citep{Philcox:encore,Slepian:candenza}. \cite{hou2022:parity} applied this idea to Sloan Digital Sky Survey Baryon Oscillation Spectroscopic Survey (SDSS BOSS) data~\citep{Dawson2013}, resulting in a $7\sigma$ detection of the parity-odd 4PCF for the larger, higher-redshift CMASS sample and a $3\sigma$ detection for the smaller, lower-redshift LOWZ sample.~\cite{philcox_parity} performed a similar study on CMASS only with the same covariance matrix \cite{Hou2022:AnalytCov} but different choices in galaxy weights and binning scheme, finding $\sim\!3\sigma$. 

A central challenge  of these analyses is how well the covariance used matches the true covariance of the data, as discussed extensively in \cite{hou2022:parity, Cahn2021:parity}. To deal with this challenge, ~\cite{Krolewski2024:parity} proposed a novel method that exploited cross-correlating  independent patches of BOSS to try to separate out any true signal from mis-estimated covariance; a similar test using the north and south galactic caps of CMASS was already performed in~\cite{hou2022:parity}, which showed that a true signal could be present yet lead to inconclusive cross-correlation. The setup of ~\cite{Krolewski2024:parity} was slightly different and found detection significances ranging from null to a maximum of $2.5\sigma$ if all the patches from the same hemisphere were combined. 

A central aspect of parity-odd 4PCF analyses is their large number of degrees of freedom, which makes the covariance an even more difficult challenge. Compressed statistics, to reduce the number of degrees of freedom, are thus another route that has been pursued. Important early work by~\cite{Jeong2012:fossil} suggested one such statistic. More recently,~\cite{Jamieson2024:POP} introduced two types of parity-odd power spectra (POP). These compressed statistics are complementary to the galaxy 4PCF, since they are sensitive to different parity-odd tetrahedral shapes (see footnote ~\footnote{The compressed statistics in~\citep{Jeong2012:fossil,Jamieson2024:POP} are sensitive to the ``squeezed'' or the ``collapsed'' types of tetrahedra. In contrast, due to the binning choices used in~\citep{hou2022:parity,philcox_parity}, the galaxy 4PCF is more sensitive to equilateral shapes. While the binning scheme for the galaxy 4PCF could be extended to probe other types, this would potentially lead to higher computational costs.}).

Given the theoretical work, observational results, and algorithmic developments, it is worth exploring additional methods for determining whether the observed parity violation has a genuine cosmological origin. One approach is simply to use a larger survey volume or different, independent tracers. Another approach is to consider whether there is a parity-violating model that fits the data (\textit{e.g.}~\cite{Niu2022:PVtk,Cabass2022:nogo, Cabass2022:Ghost, Creque-Sarbinowski2023:parity,Fujita2024:pvAxion} for recent examples). A third approach is to investigate whether there are distinctive characteristics of a true signal that might serve as a complementary test~\citep{hou2022:parity}.

Here we propose to use the Baryon Acoustic Oscillation (BAO) features in the distribution of galaxies as such a complementary test. BAO originate from a sharp feature in the baryon velocity at the moment of decoupling  ($z$$\sim$$1,020$) produced by acoustic waves in the baryon-photon plasma before that time  \cite{Sakh_66, Peebles_70, Sun_70, Hu_95}. This velocity feature, at a scale of approximately $100\,\mpch$, serves as the initial condition for the subsequent growth of the baryon density perturbations \cite{vishniac}. The dark matter, through gravitational attraction, then  converges with the baryons, resulting in the observed BAO feature at  late times, as an excess probability of finding galaxies separated by this characteristic scale~\cite{ ESW_07}. These  features  have already been detected in the galaxy 2PCF and power spectrum \citep{Eisenstein_05, Cole_05} and the galaxy 3PCF and bispectrum  \citep{SE_3PCF_BAO, Pearson_18}, as well as pointed out in theoretical modeling of  the parity-even, gravitationally-induced 4PCF \cite{ortola}.

Any pattern that existed in the density field prior to decoupling, such as parity-violating correlations produced during inflation, would carry the imprint of BAO. However, the absence of BAO in a potential parity-odd signal would also have important implications. There are two critical ``time stamps'' to consider: the end of inflation, and the moment of decoupling. The first time stamp is important because current models for the parity-odd galaxy 4PCF primarily focus on inflationary mechanisms. The second, decoupling, is the time stamp associated with BAO. {The absence of BAO in a parity-odd signal would raise significant doubts about its pre-decoupling origin, provided there is  enough signal-to-noise ratio in the overall parity-odd detection that BAO would be expected to be evident.}

The primary goal of the present work is to demonstrate the feasibility of searching for BAO imprints in the odd 4PCF. For demonstration purposes, in this work we will only consider parity-violating mechanisms of inflationary origin. 

An important additional goal of this work is to propose a model-agnostic approach for the BAO search in  the parity-odd sector; this offers the same flexibility that the original 4PCF test proposed in \cite{Cahn2021:parity}. However, this approach also presents similar challenges, such as {potential} spurious detection due to underestimated covariance ({\it e.g.} as discussed in~\cite{Cahn2021:parity,hou2022:parity}). Moreover, a simultaneous detection of both a parity-odd signal and BAO features in it does not guarantee that systematics are fully excluded. To ensure this, a specific model for parity violation would still be required.

This work is structured as follows. In \S\ref{sec:framework}, we outline the framework for understanding how BAO imprint on the density field and their effect on the trispectrum. We then briefly review the 4PCF and the isotropic basis functions. We next present a parity-violating simulation with a template whose signal peaks in the large-scale, ``soft'' (low  wavenumber) limit to show one example where we see an overall parity-odd signal and the BAO imprinted on it. In \S\ref{sec:bao_search}, we propose a model-agnostic approach for the BAO search and outline an efficient numerical method for performing it. This method remains model agnostic by essentially using the data as its own model. As we show, this ``de-wiggling'' can  actually be done at the estimator level rather than the field level, by modifying the radial binning function used in the 4PCF measurement. This makes the ``de-wiggling'' highly efficient, with the same cost as measuring a 4PCF. In \S\ref{sec:discussion}, we discuss a possible spurious significance offset introduced by this de-wiggling if applied in the absence of a real parity-odd signal. We also explore the impact of various systematics and their potential to degrade the BAO signal. Finally, we consider the prospect of using BAO from the parity-odd 4PCF as a new standard ruler. \S\ref{sec:summary} concludes.

% \noindent{Notations and Conventions}: 

\section{BAO and the Parity-Odd Galaxy 4PCF}
\label{sec:framework}

In this work, we consider a scenario where a parity-violating mechanism during inflation generates initial curvature perturbations $\calR(\bfx)$ (but see further discussion in our footnote~\footnote{While in this work our focus is on parity violation from the early Universe  (\textit{i.e.} prior to decoupling, most likely during inflation), we do not exclude the possibility that BAO could be present in the odd 4PCF if this latter is produced by some late-time, new-physics transformation of the density field. Since the density field would contain  BAO, they would remain under this late-time, new-physics, parity-violation transformation unless it somehow especially conspired to destroy them.}. The curvature perturbations then seed the density perturbations at late times, ${\delta}(\bfx) \equiv (\rho(\bfx) -\bar{\rho})/\bar{\rho}$. $\rho(\bfx)$ is the density at $\bfx$ and $\bar{\rho}$ the Universe's average density. In  Fourier space, the late-time perturbation at redshift $z$, $\tilde{\delta}(\bfk, z)$, is related to  the curvature perturbation via the matter transfer function $\calT_{\delta}(k)$ multiplied by the linear growth rate $D(z)$:
\begin{eqnarray}\label{eqn:phi_to_delta}
    \tilde{\delta}(\bfk, z) &=& -\frac{2}{5} \left(\frac{k}{\calH}\right)^2 \frac{D(z)}{D_0}  \calT_{\delta}(k) \tilde{\calR}(\bfk),
\end{eqnarray}
with $D_0\equiv D(z=0)$, $\calH = aH$ the comoving Hubble parameter, $H$ the Hubble parameter, and $a$ the scale factor (we discuss the relationship to the potential, $\phi$, in footnote \footnote{During the radiation-dominated era, the density perturbation and the potential $\phi$ are related via $\delta = -(2/3) [k/(aH)]^2 \phi$. In this work, we employ the curvature perturbation, which is related to the primordial potential as ${\calR} = 5/3 \, {\phi}$, hence the pre-factor of $-2/5$ above, while the minus sign stems from the Poisson equation, from which  the factor of $(k/\calH)^2$ also arises.}). Tilde denotes a field in Fourier space. Importantly, the matter transfer function encodes the BAO.

As noted in \S\ref{sec:intro}, the lowest-order statistic sensitive to parity in the density field is the 4PCF \cite{Cahn2021:parity}; here, we begin with its Fourier space analog, the trispectrum. The primordial curvature trispectrum $T_{\calR}(\bfk_1,\bfk_2, \bfk_3,\bfk_4)$ is defined via
\begin{align}
\av{\prod_{i=1}^4\tilde{\calR}(\bfk_i)
    }\equiv
    (2\pi)^3\delta^{[3]}_{\rm D}(\bfk_{1234})\, T_{\calR}(\bfk_1,\bfk_2, \bfk_3,\bfk_4),
\end{align}
where $\tilde{\calR}(\bfk_i)$ is the primordial curvature perturbation in Fourier space, $\delta_{\rm D}^{[3]}(\bfk_{1234})$ is a 3D Dirac delta function and $\bfk_{1234}\equiv \sum_{i=1}^4 \bfk_i$. 

The primordial trispectrum $T_{\calR}$ can be split into a parity-even component $\calT_{\calR, +}$ and a parity-odd component $\calT_{\calR, -}$. We have  
\begin{align}
\label{eqn:trispec_even_odd}
    T_{\calR}(\bfk_1,\bfk_2, \bfk_3,\bfk_4) &= \mathcal{T}_{\calR,\, +}(\bfk_1,\bfk_2,
    \bfk_3,\bfk_4) \nonumber\\ &+ i\,\mathcal{T}_{\calR,\, -}(\bfk_1,\bfk_2,
    \bfk_3,\bfk_4).
\end{align}
The imaginary, parity-odd part will be non-zero on large scales only if there is beyond-Standard Model physics during inflation.

In linear theory, the primordial trispectrum $T_{\calR}$ is evolved into the matter trispectrum $\mathcal{T}_{\rm m}$ by applying the matter transfer functions, leading to
\begin{align}
\label{eqn:trispec_pri_gal}
   &\mathcal{T}_{\rm m}(\bfk_1,\bfk_2, \bfk_3,\bfk_4)\\
   &= \prod_{i=1}^4 \left[\frac{2}{5} \left(\frac{k_i}{\calH}\right)^2 \frac{D(z)}{D_0} \calT_{\delta}(k_i)\right] T_{\calR}(\bfk_1,\bfk_2, \bfk_3,\bfk_4).\nonumber
\end{align}
Everything inside the square brackets has a product taken over it; in other words even the un-subscripted pre-factors should be repeated.

Now, since the right-hand side of Eq. (\ref{eqn:trispec_pri_gal}) has real and imaginary parts, so will the left-hand side. In short, the late-time linear matter trispectrum will have even and odd pieces if the primordial trispectrum does. We note that both even and odd parts of the linear matter trispectrum will be subject to the standard non-linear gravitational evolution \textit{e.g.}~\cite{bernardeau}.

Combining Eqs.~\eqref{eqn:trispec_even_odd} and ~\eqref{eqn:trispec_pri_gal}, we see that BAO imprint on the parity-odd late-time trispectrum through the product of transfer functions $\calT_{\delta}$. Thus, if there is a non-zero parity-odd primordial trispectrum, the BAO will imprint on it, and hence manifest in the parity-odd linear matter trispectrum and, eventually, in the parity-odd galaxy trispectrum.

Finally, the linear matter 4PCF, $\zeta$, is the inverse Fourier Transform (FT) of the linear matter trispectrum $\mathcal{T}_{\rm m}$:
\begin{align}
\label{eqn:4PCF_ift}
    &\zeta(\bfr_1,\bfr_2,\bfr_3) = \\ 
    &\qquad\mathcal{FT}^{-1} \left[(2\pi)^3\delta^{[3]}_{\rm D}(\bfk_{1234})\, \mathcal{T}_{\rm m}(\bfk_1,\bfk_2, \bfk_3,\bfk_4)\right],\nonumber
\end{align}
where $\bfr_i\equiv \bfx_i-\bfx_0$ for $i = 1, 2, 3$. $\bfx_i$ and $\bfx_0$ are the absolute positions of four points, but due to the cosmological assumption of homogeneity, we have translation symmetry, so we may losslessly average over $\bfx_0$. This averaging is equivalent by ergodicity to averaging over realizations of the density field, and enables eliminating $\bfx_0$ in favor of the relative coordinates $\bfr_i$. The Dirac delta $\delta^{[3]}_{\rm D}(\bfk_{1234})$ in Fourier space can be understood as a consequence of this averaging; the Delta function enforces momentum conservation, and momentum is the conserved quantity associated with translation invariance.

Here and  throughout, our FT convention is that $\FT{f(\bfx)} \equiv \int d^3\bfx \, e^{-i\bfk\cdot \bfx} f(\bfx)$ and $\iFT{\tilde{f}(\bfk)}\equiv\int_{\bfk} e^{i\bfk\cdot \bfx} \tilde{f}(\bfk)$, where $\int_{\bfk}\equiv \int d^3\bfk/(2\pi)^3$. 

\section{Illustration of BAO in the 4PCF}
In this section, we illustrate the BAO signal in the parity-odd 4PCF. We will first review the 4PCF in the isotropic basis of \cite{Cahn202010}. Then we will compute an example detection significance of the BAO signal using a suite of simulations with a toy model parity-violating 4PCF encoded in them.

\subsection{4PCF in the Isotropic Basis}
A direct measurement of the 4PCF in Eq.~\eqref{eqn:4PCF_ift} is computationally challenging. To accelerate the process, we decompose the 4PCF into a basis of isotropic basis functions $\mathcal{P}_{\ell_1 \ell_2 \ell_3}$ \cite{Cahn202010} that capture its angular behavior about one galaxy (the ``primary';), times radial coefficients that capture its dependence on tetrahedron side lengths $r_1, r_2, r_3$ from that primary. The isotropic basis~\citep{Cahn202010} has rotational invariance and can well capture the isotropic information in the large-scale distribution of galaxies. The radial coefficients are
\begin{align}
    &\zeta_{\ell_1\ell_2\ell_3}(r_1,r_2,r_3) = \\
    &\int d\hr_1\,d\hr_2\,d\hr_3\, {\zeta}(\bfr_1,\bfr_2,\bfr_3) \,\mathcal{P}^*_{\ell_1 \ell_2 \ell_3}(\hat{r}_1, \hat{r}_2, \hat{r}_3),\nonumber
\end{align}
where $\ell_i$, for $i=1,2 ,3$ are the angular momenta associated with the three direction vectors $\bfr_i$,  and star denotes a conjugate. 

To avoid an over-complete basis, we assign physically observable meaning  to the subscripts $1,2,3$ by ordering the tetrahedron sides so that  $r_1 < r_2  < r_3$. The isotropic functions $\mathcal{P}_{\ell_1 \ell_2 \ell_3}(\hat{r}_1, \hat{r}_2, \hat{r}_3)$ of three arguments are: 
\begin{align}
\label{eqn:Plll}
\mathcal{P}_{\ell_1 \ell_2 \ell_3}(\hat{r}_1, \hat{r}_2, \hat{r}_3) &= \sum_{m_1 m_2 m_3} 
(-1)^{\ell_1+\ell_2+\ell_3}\six{\ell_1}{\ell_2}{\ell_3}{m_1}{m_2}{m_3}\nonumber\\
&\quad \times Y_{\ell_1 m_1}(\hat{r}_1)
Y_{\ell_2 m_2}(\hat{r}_2)
Y_{\ell_3 m_3}(\hat{r}_3),
\end{align}
where the $Y_{\ell_i m_i}(\hr_i)$ are spherical harmonics and the $m_i$ are the $z$-components of their angular momenta $\ell_i$; the $m_i$ are also called the ``projective quantum number''. The isotropic basis is fully separable in the $\bfr_i$, thus reducing the formal complexity of the 4PCF estimator to pair-wise operations \cite{SE_3pt, Philcox:encore} (the actual scaling in practice is outlined in our footnote \footnote{In detail, the actual execution time for typical spectroscopic samples (such as BOSS) ends up scaling linearly in the number of objects because computation of the harmonic expansion coefficients is sub-dominant to the cost of assembling all triples of them, which must be done around each galaxy and so scales linearly.}). 

While the full trispectrum in Eq.~\eqref{eqn:trispec_even_odd} encodes both even and odd parts, it is non-trivial to isolate only the odd (imaginary) part of the trispectrum in practice. From Eq.~\eqref{eqn:Plll} we see that the basis functions behave under parity, $\mathbb{P}$: $(x,y,z) \rightarrow (-x,-y,-z)$, as
\begin{eqnarray}
    &&\mathbb{P}\left[\mathcal{P}_{\ell_1 \ell_2 \ell_3}(\hat{r}_1, \hat{r}_2, \hat{r}_3)\right]=\mathcal{P}_{\ell_1 \ell_2 \ell_3}(-\hat{r}_1, -\hat{r}_2, -\hat{r}_3)
    \nonumber\\ &&\qquad =(-1)^{\ell_1+\ell_2+\ell_3}\mathcal{P}_{\ell_1 \ell_2 \ell_3}(\hat{r}_1, \hat{r}_2, \hat{r}_3).
\end{eqnarray}
An odd sum of $\ell_i$ therefore means a parity-odd  basis function. The odd functions pick up any difference between the frequency of appearance of a given tetrahedron, and of its mirror image, in a galaxy sample. Thus, if  they are non-zero at a statistically significant level, they reveal parity violation in the sample. 

We note that, in the presence of realistic survey geometry, one must apply a mode decoupling matrix to correct for the mode mixing due to the geometry's breaking the orthogonality of the basis functions~\cite{SE_3pt, Philcox:encore}. We return to this point in \S\ref{sec:dewiggling}. 

\subsection{Detection Significance of BAO in Parity-Odd Toy Simulations}

\subsubsection{Setup and Simulations}
\label{subsec:setup_sim}
In this section, we discuss the setup of the simulations that we use to illustrate the BAO imprint on the odd 4PCF. As discussed in \S\ref{sec:framework}, we consider the parity-violating mechanism to arise from some inflationary scenario. Accordingly, we simulate the curvature perturbation \(\calR(\bfk)\), which includes both a Gaussian (subscript ``G'') and a non-Gaussian (subscript ``NG'') component:
\begin{align}
    \calR(\bfk) = \calR_{\rm G}(\bfk) + \calR_{\rm NG} (\bfk).
\end{align}
A simple template for $\calR_{\rm NG}$ that produces an odd trispectrum is~\citep{Coulton2023:parity,Jamieson2024:POP}
\begin{align}
\calR_{\rm NG} (\bfk) &= ig \int_{\bfq_1\bfq_2\bfq_3} \delta^{[3]}_{\rm D}(\bfk-\bfq_{123})\, \frac{\bfq_1\cdot(\bfq_2\times\bfq_3)}{q_1^{\alpha}q_2^{\beta} q_3^{\gamma}} \nonumber\\
&\,\,\times \calR_{\rm G}(\bfq_1)\calR_{\rm G}(\bfq_2)\calR_{\rm G}(\bfq_3),
\end{align}
with $\bfq_{123}\equiv \bfq_1+\bfq_2+\bfq_3$, and $g$ a coupling constant that controls the amplitude of the non-Gaussianity. $\alpha, \beta,$ and $\gamma$ are free parameters that satisfy $\alpha+\beta+\gamma=-3$ to keep $g$ dimensionless. In particular, we choose $\{\alpha, \beta, \gamma\} = \{-2,-1,0\}$, following~\citep{Coulton2023:parity,Jamieson2024:POP}.

The leading-order parity-odd primordial curvature trispectrum from this template is
\begin{eqnarray}
\label{eqn:trispec_template}
&&T_{-}(\bfk_1, \bfk_2, \bfk_3, \bfk_4)\\
&=& g \, \bfk_1 \cdot (\bfk_2 \times \bfk_3) \left(2 \pi^2 A_\mathrm{s}\right)^3 \nonumber\\
&&\times\,\left(k_1^{\alpha-4+n_{\rm s}} k_2^{\beta-4+n_{\rm s}} k_3^{\gamma-4+n_{\rm s}} k_4^{0} \pm \mathrm{23\ signed\ perms.}\right), \nonumber
\end{eqnarray}
where ``perms.'' means to permute the four wave numbers $\{k_1,k_2,k_3,k_4\}$ relative to the powers $\alpha, \beta, \gamma, 0$. There are 23 permutations of the original ordering, and the sign is determined by whether it is cyclic ($+1$) or anti-cyclic ($-1$). 

Although this template is not directly linked to any particular physical model, it is useful for gaining insight into how BAO might manifest within an odd 4PCF. We note that the template is fully separable in the $\bfk_i$. By isolating the scalar triple product $\bfk_1 \cdot (\bfk_2 \times \bfk_3)$, the remaining parts are parity-even and have only radial dependence. A previous analysis~\citep{Jamieson2024:POP} revealed that this template peaks in the ``soft'' (low-$k$) limit. Specifically, the template reaches its maximum when $k_1 \rightarrow 0$ with the wavenumbers ordered as $k_1 < k_2 < k_3 < k_4$. However, the range of scales explored in the galaxy 4PCF thus far has been restricted to tetrahedra where the sides are more roughly equal in length~\citep{hou2022:parity,philcox_parity}, making the galaxy 4PCF (at least as used thus far) less sensitive to this `squeezed' trispectrum template. To increase sensitivity, we could study the signal over a much larger range of scales (which might  pose algorithmic challenges). Alternatively, we could consider other templates that have higher amplitude in roughly equilateral 4PCF configurations. We leave this for future work.

We construct simulations with both positive and negative coupling constants, $g = \pm 2\times 10^7$ (see our footnote \footnote{This choice of $|g|=2\times 10^7$ approaches the upper limit of the coupling constant. Selecting a much higher value would introduce higher-order contributions, such as $\av{\calR \calR \calR^{(3)} \calR^{(3)}}$, which would cause the likelihood to deviate significantly from a Gaussian distribution.}), calibrated so that their power spectra match that of the simulation without a non-Gaussian field. We then apply transfer functions with and without BAO features to these density fields, and evolve to redshift $z=0$ using \textsc{CLASS}~\citep{Lesgourgues2011:class}. The simulations with positive and negative coupling constants, as well as those with and without BAO features, all share the same initial conditions. In total, there are $100\times 4=400$ simulation boxes, each with $N_{\rm mesh}=256$ mesh points per side. We interpret the box side length to be $L_{\rm box}=1\,\gpch$, which corresponds to a cell resolution of $\Delta L = 3.91 \, \mpch$. Following the analysis of~\citep{hou2022:parity}, we measure the 4PCF with a minimum tetrahedron side length of $r_{\rm min}=20 \,\mpch$ and a maximum value of the $r_i$ of $r_{\rm max}=160 \,\mpch$, with a radial bin width of $\Delta r=14\,\mpch$. We note that this maximum is on any side extending  from the ``primary'' point; a side that does not attach to the primary can range  up to twice this (from having an angle of 180 degrees between two sides that do). Information about sides that do not attach to the  primary is captured in the $\ell_i$ dependence of the 4PCF coefficients. 

\begin{figure*}
    \includegraphics[width=0.8\linewidth]{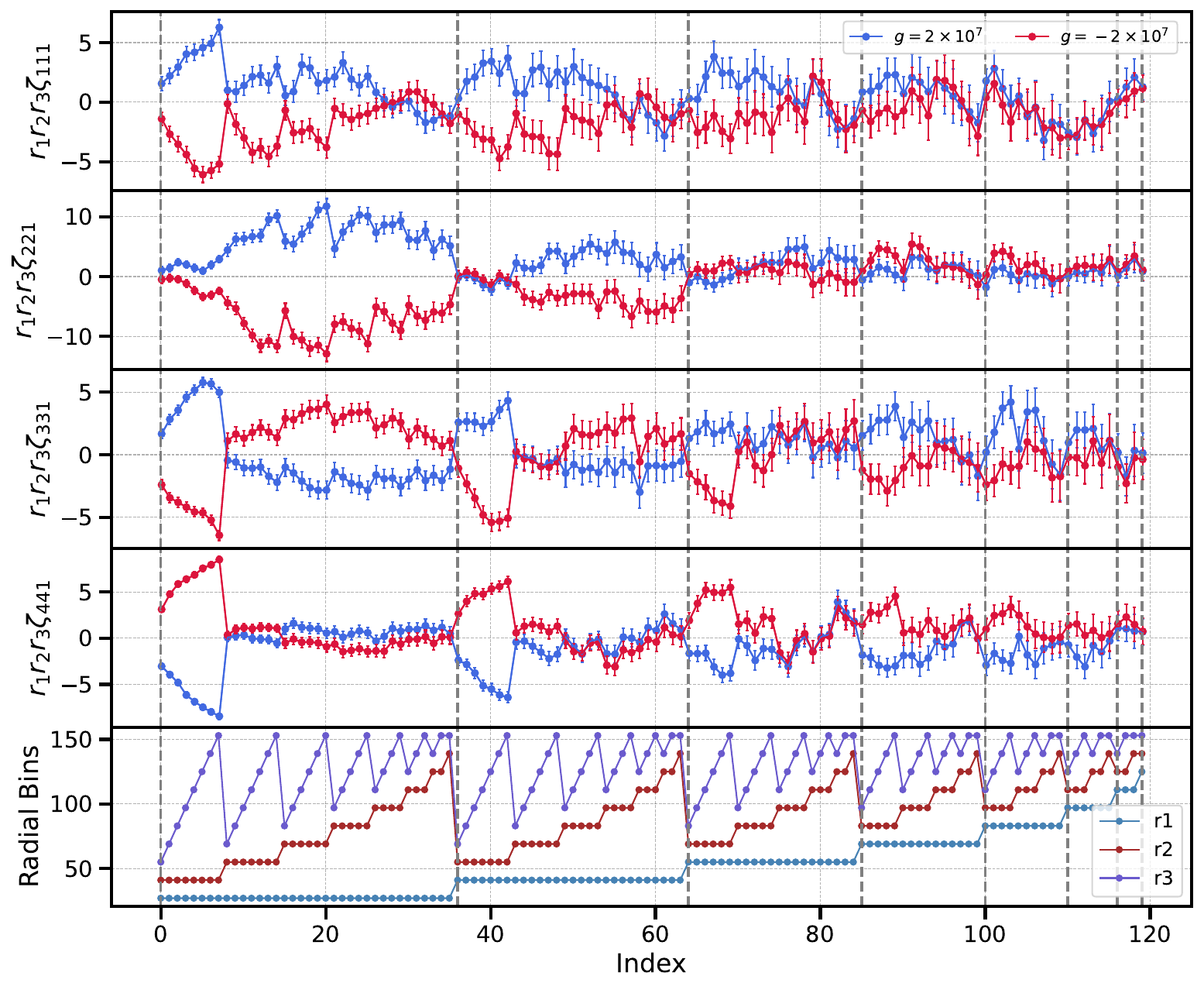}        
    \caption{Measurement of the lowest-lying few 4PCF coefficients on parity-violating simulations with coupling constants $g=\pm 2\times 10^7$, (the template is in Eq. \eqref{eqn:trispec_template}). For each $g$ we have 100 averaged realizations. Here we show the angular channels for $\ell_1=\ell_2\leq 4$ at a fixed $\ell_3=1$. As explained in \cite{Cahn202010} one must order the side lengths such that $r_1 < r_2 < r_3$ of the tetrahedra to avoid redundancy of the basis. For plotting we mapped them to a 1D index, and this mapping is displayed in the bottom panel. Since the template peaks in the squeezed limit, the signal-to-noise ratio is higher when the three sides of the tetrahedron are far apart and lower when the sides are of similar length. To guide the eye, we emphasize the transition in the $r_1$ index.}
    \label{fig:toy_pv_256_4PCF_odd_20x160x10_ll1}
\end{figure*}
Fig.~\ref{fig:toy_pv_256_4PCF_odd_20x160x10_ll1} shows the measurements of the 4PCF coefficients for angular channels $\ell_1=\ell_2\leq 4$ at a fixed $\ell_3=1$ (both odd parity); the lengths and arrangement of the three sides $\{r_1, r_2, r_3\}$ is shown in the bottom panel. We notice that the signal is maximized when $\ell_3=1$ and when the separations between $r_1$, $r_2$, and $r_3$ are the largest. This behavior is consistent with our Fourier-space template's peaking in the ``soft'' limit, which in position space means when the tetrahedron side lengths are most disparate. As the three sides approach equal size (equilateral), the signal significantly drops. As shown in Fig.~\ref{fig:toy_pv_256_4PCF_odd_20x160x10_ll1}, starting from index $\sim\! 80$, the signal-to-noise ratio declines, making it difficult to distinguish between the positive and negative coupling constants.

\begin{figure*}
    \centering
    \begin{subfigure}[b]{0.99\textwidth}
         \centering
    \includegraphics[width=.95\textwidth]{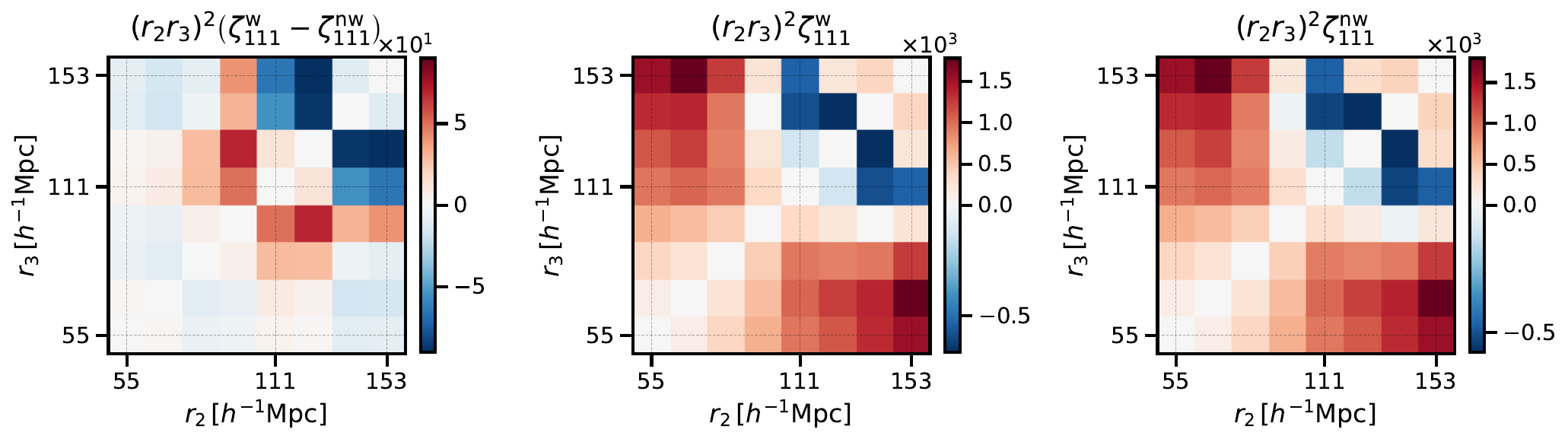}
        \label{fig:toy_pv_256_20x160xnbin10_rfix_at_41_lmax5_4PCF2d_odd_ell111}
     \end{subfigure}
    \begin{subfigure}[b]{0.99\textwidth}
         \centering
    \includegraphics[width=.9\textwidth]{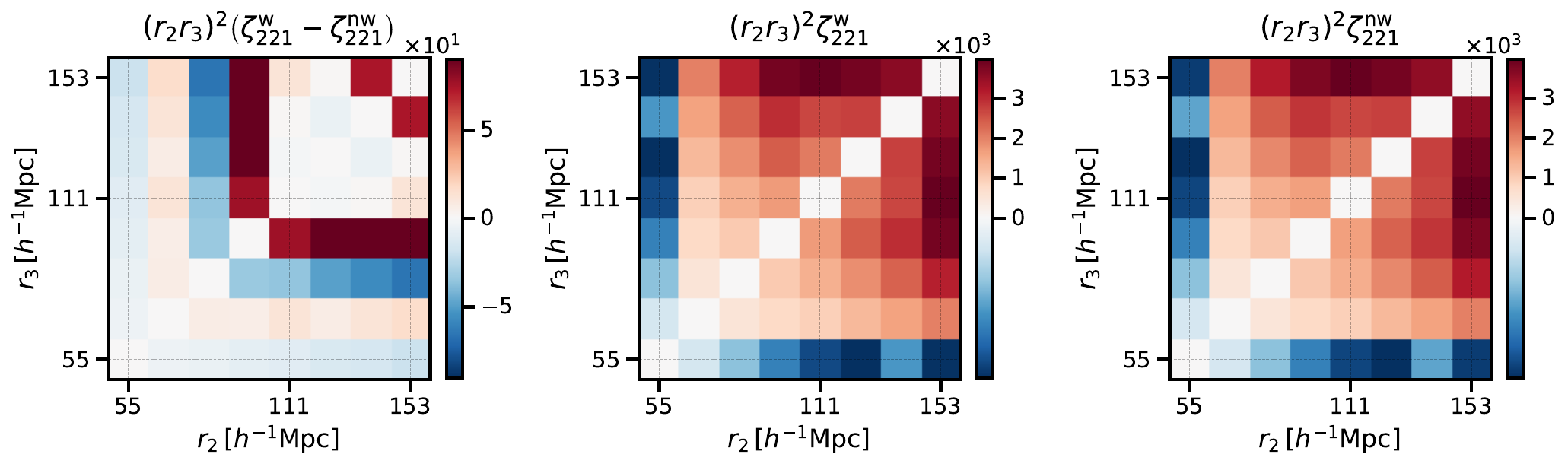}
         \label{fig:toy_pv_256_20x160xnbin10_rfix_at_41_lmax5_4PCF2d_odd_ell221}
     \end{subfigure}    
    \caption{Here we show the BAO features in a few low-lying angular channels of the parity-odd 4PCF, $\zeta_{\ell_1\ell_2\ell_3}(r_2, r_3)$, vs. the middle and longest tetrahedron side lengths, $r_2$ and $r_3$, fixing the shortest, $r_1$, to be $41\,\mpch$. The left-hand panel shows the difference between the 4PCF coefficients with BAO and without BAO. The middle panel shows the 4PCF without BAO, and the right panel shows it with BAO. Thus the leftmost panel is the difference of the middle and the right-most. The top row shows the $\ell_1, \ell_2, \ell_3$ = $1,1,1,$ channel and the bottom row shows the $\ell_1, \ell_2, \ell_3$ = $2,2,1,$ channel. These plots are all weighted by $(r_2 r_3)^2$ since the uniformly random distributed galaxy counts in each radial bin scale as $r_i^2$.}
\label{fig:toy_pv_256_20x160xnbin10_rfix_at_41_lmax5_4PCF2d_odd_ell}
\end{figure*}
{Fig.~\ref{fig:toy_pv_256_20x160xnbin10_rfix_at_41_lmax5_4PCF2d_odd_ell} shows the parity-odd 4PCF measurements for two angular channels, $\{\ell_1,\ell_2,\ell_3\}=\{1,1,1\}$ and $\{\ell_1,\ell_2,\ell_3\}=\{2,2,1\}$, as functions of two of the side lengths ($r_2$ and $r_3$), with the shortest side ($r_1$) held fixed. 
The left-most panels show the difference between the parity-odd 4PCF coefficients with (superscript ``w''), and without (superscript ``nw'') the BAO, at fixed $r_1=41\,\mpch$. The middle and right-most panels are the original 4PCF measurements respectively with and without BAO. Although the BAO is only a few per cent effect, and the original 4PCF coefficients appear nearly identical, the difference plots reveal that the BAO manifest as transition from red to blue (excess of configurations over and above random, to decrement) or blue to red (vice versa) at scales of approximately $\sim\! 100\, \mpch$. 

\subsubsection{Overall Parity-Violation Detection Significance}
\label{sec:significance}
We first compute $\chi^2$ for the \textit{overall} detection significance of the parity-odd signal:
\begin{eqnarray}
\label{eqn:chi-squared-twelve}
    \chi^2 = \sum (\zeta_{\rm d}-\zeta_{\rm m}) \mathbb{C}^{-1} (\zeta_{\rm d}-\zeta_{\rm m})^{\rm T}.
\end{eqnarray}
$\mathbb{C}$ is the covariance matrix, $\zeta_{\rm d}$ is the measured 4PCF from the data (hence, subscript ``d''), and $\zeta_{\rm m}$ is a 4PCF model (hence subscrript ``m''). Following~\citep{hou2022:parity}, we take our null hypothesis to be that $\zeta_{\rm m} = 0$.

We consider two methods of computing the covariance matrix. The first uses the analytic template of ~\citep{Hou2022:AnalytCov}, which takes the density fluctuation field to be Gaussian Random. The second uses 5,000 Gaussian Random Fields that we constructed to have matching power spectrum to that of the parity-violating simulations. In this second method, the $\chi^2$ distribution is modified to be a multivariate $T$-distribution~\citep{Sellentin2016:T2} to account for noise in the covariance since it is estimated from a finite sample. We further explore the impact of covariance matrix method choice in Appendix~\ref{appendix:cov}, and show it has only a marginal impact on the significances (though see footnote \footnote{We caution that the choice of binning can impact the significances). Using bins that are too fine may result in a non-Gaussian likelihood because one cannot invoke central limit theorem to argue that the 4PCF coefficient at each radial bin combination is Gaussian-distributed. In this case using the analytic covariance may not result in the expected $\chi^2$ distribution.}.

Fig.~\ref{fig:detxn_4pcf_odd_20x160x10} shows the detection significance of the overall parity-odd signal in the lefthand panel with both the positive and negative coupling constants. Here, we treat each simulation as one realization of the Universe. We then  quantify the detection significance for each realization in units of the standard deviation of the null distribution, given by our Gaussian simulations. The overall detection significance for parity violation in each realization is typically about $4\sigma$.

\subsubsection{BAO Detection Significance}
\label{subsub:bao-sig}
Now, to assess the \textit{BAO} detection significance, we need to develop an alternative to the standard approach used in BAO searches for 2- and 3-point statistics ({\it e.g.}~\cite{Eisenstein_05,Cole_05,SE_3PCF_BAO,Pearson_18}). The standard approach computes the difference between the $\chi^2$ of the observable under the best-fit model with BAO, and the $\chi^2$ of the observable under the best-fit model with no BAO. This difference is $\Delta \chi^2$, and $\sqrt{\Delta\chi^2}$ between the two models is the BAO detection significance. The presence in the $\chi^2$ of the inverse covariance matrix, $\mathbb{C}^{-1} = \mathbb{S}^{-1} \mathbb{D}^{-1} \mathbb{S}$ both rotates the data vectors into an orthogonal basis (role of $\mathbb{S}$ and $\mathbb{S}^{-1}$), where each mode is independent, and also scales each such mode to have variance of unity (role of $\mathbb{D}^{-1}$). Thus the $\sqrt{\Delta \chi^2}$ is simply the distance between the two models in units suitable to be compared with a normal distribution (mean zero, unit variance) when computing probabilities.

However, the standard approach requires a model for the expected signal, which is not known {\it a priori}. Instead, we propose a different method that uses the data as its own model. We can remove the BAO from the density field and use the ``de-wiggled'' observable (4PCF measured on the de-wiggled density field) in place of the ``no wiggle'' model employed in the standard approach. We denote the $\chi^2$ difference computed in this fashion with a tilde.

Our $\chi^2$ difference  is now
\begin{eqnarray}\label{eqn:Delta_tilde_chi2_bao}
    \Delta\tilde{\chi}^2_{\rm bao} = (\zeta_{\rm d} - \zeta^{\rm nw}_{\rm d}) \mathbb{C}^{-1} (\zeta_{\rm d} - \zeta^{\rm nw}_{\rm d})^{\rm T},
\end{eqnarray}
where $\zeta_{\rm d}$ and $\zeta^{\rm nw}_{\rm d}$ are the 4PCF measured from the same data, but superscript ``nw'' means  de-wiggling has been applied first. In Appendix~\ref{appendix:BAO_detxn}, we show that in the high signal-to-noise ratio limit, our new statistic $\Delta\tilde{\chi}^2_{\rm bao}$ recovers the mean of the usual $\Delta \chi^2_{\rm bao}$. In the low signal-to-noise ratio limit, the mean of the new statistic is more sensitive to statistical noise, particularly for high-dimensional data vectors. Nevertheless, since the both $\zeta$ are measured from the same data, they also have the comparable noise (up to the effect of de-wiggling), the variance in the new statistic is also reduced.

 It is important to notice that the ``de-wiggling'' approach proposed here does result in an offset in the modified statistic $\Delta\tilde{\chi}^2_{\rm bao}$. In particular, even in the absence of a true parity-odd 4PCF, the expected BAO detection significance $\Delta\chi^2_{\rm s}$ is non-zero. This occurs because the de-wiggling process itself alters the spatial configuration of the density field, regardless of whether there is any true parity-odd component.
When we compute the $\Delta \chi^2_{\rm bao}$, in the absence of a true parity-odd 4PCF, we are just comparing the variance of two data vectors (with BAO, and without  BAO) that should each be centered around zero. However, the presence or absence of BAO causes the variances to differ, leading to a non-vanishing $\Delta \chi^2_{\rm bao}$. This difference ultimately can be understood by examining the ratio of the matter transfer function with BAO to that without BAO. We return to this issue in \S\ref{sec:detxn_null}. 

To address this issue, we quantify the BAO significance by comparing $\Delta\tilde{\chi}^2_{\rm bao}$ measured from parity-violating simulations to a null distribution $\Delta\tilde{\chi}^2_{\rm bao, null}$ measured from purely Gaussian (no parity-violation) simulations. These latter already contain the variance effect outlined above, and so using them to define our null distribution removes any spurious $\Delta \chi^2$ so produced. We thus define our BAO significance as
\begin{eqnarray}
    S_{\rm bao} = \frac{\Delta\tilde{\chi}^2_{\rm bao} - \Delta\tilde{\chi}^2_{\rm bao, null}}{ \sigma \left(\Delta\tilde{\chi}^2_{\rm bao, null}\right)},
\end{eqnarray}
where $\sigma \left(\Delta\tilde{\chi}^2_{\rm bao, null}\right)$ is the standard deviation of the null distribution.

The righthand panel of Fig.~\ref{fig:detxn_4pcf_odd_20x160x10} shows the detection significance of BAO in the odd 4PCF. We treat each simulation without BAO (``nw'') as the ``model''. We then compare the $\Delta\tilde{\chi}^2_{\rm bao}(g=\pm 2\times 10^7)$ for parity-violating simulations to that for purely Gaussian simulations. We find the typical BAO significance computed in this way is about $3\sigma$. 

We pause to highlight a fundamental difference in the detection significance quoted here as compared to that used in the standard BAO search method. In the standard approach, the significance is based on the signal-to-noise ratio. Here, our null hypothesis is that we have BAO but in a parity-conserving scenario; our significance reflects how strongly we may reject this null hypothesis.

\subsubsection{Dependence on Template Choice and Coupling Constant}
The actual values of the significance found here are influenced both by the template choice and the coupling constant. Given that our template is strongest in the ``soft'' (low-$k$) limit, as noted in \S\ref{subsec:setup_sim}, the detection here may well be a conservative estimate of what could be found if parity violation is genuine. This is because we might expect that the BAO signal is strongest when one side is at the BAO scale, or two sides add up to the BAO scale, or add to twice the BAO scale; this is the behavior seen in the 3PCF (for instance, figure 8, $\ell = 1$ panel, in \cite{se_rsd_3pcf}). Thus, equilateral or semi-equilateral configurations are likely preferred for BAO. Yet, these configurations do not appear in position space with very large amplitude given our template.

In the current work, we set our coupling constant $g$ to be as large as possible without introducing beyond-leading-order terms in the parity-odd trispectrum, while also ensuring that the power spectrum of the non-Gaussian simulations remains consistent with observational constraints. 

\subsubsection{Caveats}
In the real Universe, non-linear gravitational evolution affects the sharpness of the BAO feature, as is well-known for the two-point functions, and can also impact the detection significance in the odd 4PCF. We leave the impact of non-linear gravitational evolution to future work, such as using N-body codes to evolve parity-violating simulations. Furthermore, our work here assumes linear bias only, and treats the continuous simulation as a proxy for the galaxy field, when in fact the latter are discrete, which leads to Poisson noise that also will impact the BAO significance.

\begin{figure*}
    \centering
    \begin{subfigure}[b]{0.45\textwidth}
         \centering
    \includegraphics[width=.9\textwidth]{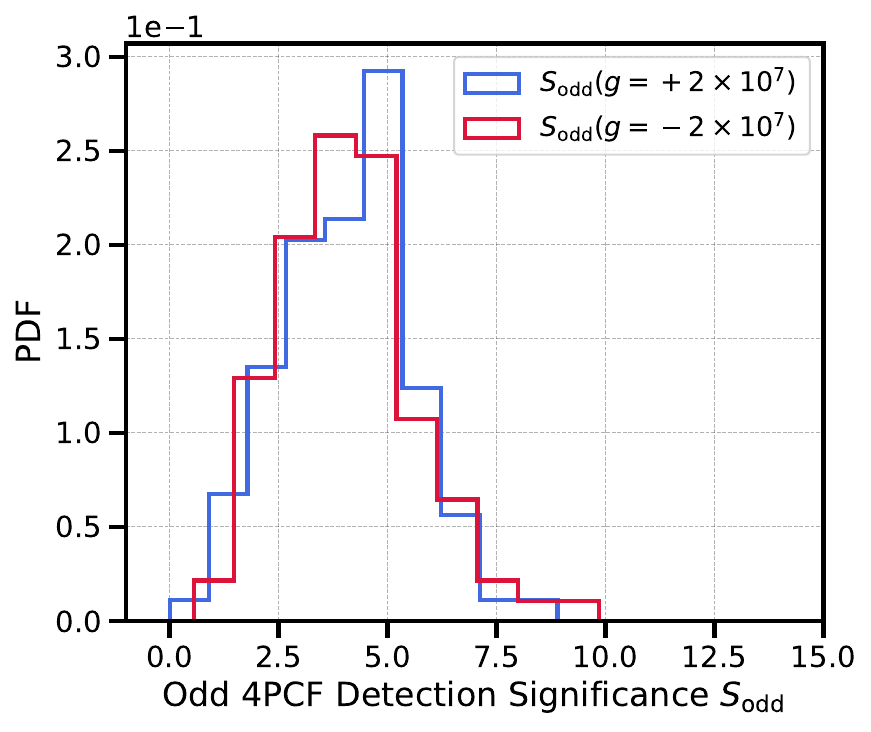}
    \end{subfigure}
    % \hfill
    \begin{subfigure}[b]{0.45\textwidth}
         \centering
    \includegraphics[width=.9\textwidth]{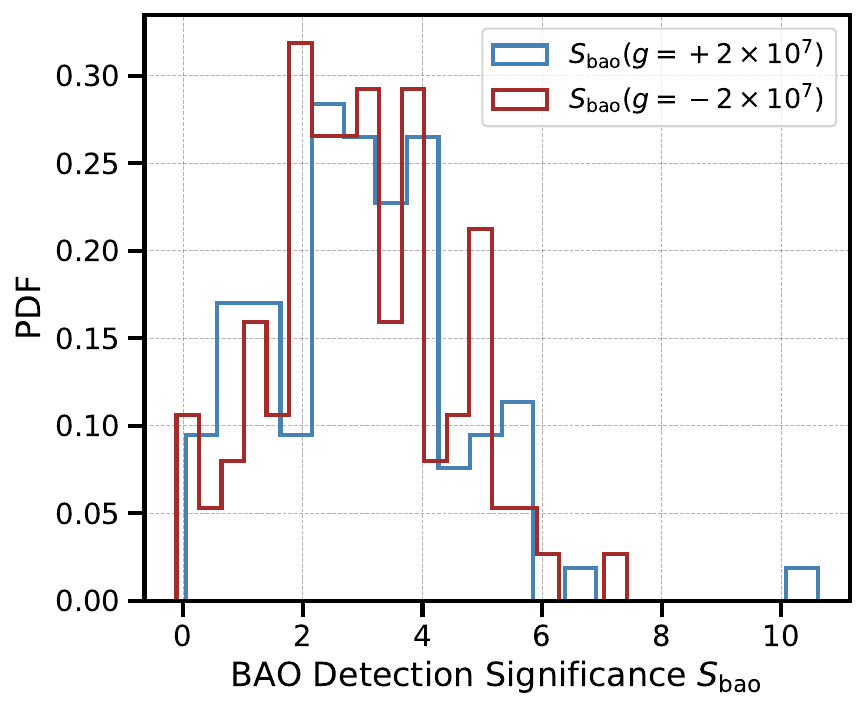}
    \end{subfigure}
    \caption{Detection significance for both coupling constants $g=\pm 2\times 10^7$. {\it Left}: Overall parity-odd detection significance. {\it Right}: BAO detection significance in the parity-odd measurement. In both measurements, we treat each simulation as an independent realization of the Universe, and quantify the excess probability in units of the standard deviation of a null distribution given by our Gaussian simulations. The significance in both cases reflects how well we can reject the null hypothesis.}
    \label{fig:detxn_4pcf_odd_20x160x10}
\end{figure*}

\section{Method for Efficient De-Wiggling and BAO Search on Observational Data}
\label{sec:bao_search}

\label{sec:dewiggling}
The 4PCF of the de-wiggled density field, $\zeta^{\rm nw}$, is one key ingredient in our BAO search ({\it cf.} Eq.~\ref{eqn:Delta_tilde_chi2_bao}).
In \S\ref{subsec:setup_sim}, we constructed simulations without BAO by evolving the primordial curvature perturbations using the no-wiggle transfer function. 

For observational data, one approach to remove the BAO feature would be to de-wiggle the density field and then measure the 4PCF. However, it would be more efficient to do this directly by slightly altering the 4PCF estimator, so that de-wiggling is done ``on the fly'' as we measure the 4PCF. Here we outline such an approach.  We begin with a brief review of the 4PCF estimator.

\subsubsection{4PCF Estimator}
The 4PCF estimator for discrete data (such as galaxies), which we  denote $\hat\zeta$, comes from generalizing \cite{LS93, SS98, SE_3pt}. We define $N(\bfx) \equiv D(\bfx) - R(\bfx)$, with $D(\bfx)$ the data and $R(\bfx)$ the randoms at a point $\bfx$ (see footnote \footnote{Random catalogs are required to convert the galaxy counts into a density fluctuation field and to correct for survey geometry}). We then have the estimator
 \begin{align}\label{eqn:4pcf_estimator}
\hat{\zeta}(\bfr_1,\bfr_2,\bfr_3) &= \frac{\int d^3\vec{x}\, {N(\vec{x}) N(\vec{x} +\vec{r}_1) N(\vec{x} + \vec{r}_2) N(\vec{x} + \vec{r}_3 )}}{ \int d^3\vec{x}\, R(\vec{x}) R(\vec{x} +\vec{r}_1) R(\vec{x} + \vec{r}_2) R(\vec{x} + \vec{r}_3 )}\nonumber\\
&\equiv \frac{\calN(\bfr_1,\bfr_2,\bfr_3)}{\calR(\bfr_1,\bfr_2,\bfr_3)},
\end{align}
where the bottom equivalence defines $\calN$ and $\calR$. Following \cite{SE_3pt} \S4 and \cite{Philcox:encore},  we may expand the 4PCF as well as the $N(\bfx)$ and the $R(\bfx)$ fields in the isotropic basis \citep{Cahn202010, Cahn2021:parity} as
\begin{align}\label{eqn:NPCF_expansion}
    &\sum_{\ell_1\ell_2\ell_3} \hat{\zeta}_{\ell_1\ell_2\ell_3}(r_1,r_2,r_3)\calP_{\ell_1\ell_2\ell_3}(\hat{r}_1, \hat{r}_2, \hat{r}_3) \nonumber\\
    &= \frac{\sum_{\lambda_1 \lambda_2 \lambda_3} \calN_{\lambda_1 \lambda_2 \lambda_3}(r_1,r_2,r_3) \calP_{\lambda_1 \lambda_2 \lambda_3}(\hat{r}_1, \hat{r}_2, \hat{r}_3)}{\sum_{\tilde{\lambda}_1 \tilde{\lambda}_2 \tilde{\lambda}_3} \calR_{\tilde{\lambda}_1 \tilde{\lambda}_2 \tilde{\lambda}_3}(r_1,r_2,r_3) \calP_{\tilde{\lambda}_1 \tilde{\lambda}_2 \tilde{\lambda}_3}(\hat{r}_1, \hat{r}_2, \hat{r}_3)},
\end{align}
where $\hat{\zeta}_{\ell_1\ell_2\ell_3}$, $\calN_{\lambda_1 \lambda_2 \lambda_3}$, and $\calR_{\tilde{\lambda}_1 \tilde{\lambda}_2 \tilde{\lambda}_3}$ are the radial expansion coefficients.

From Eq.~\eqref{eqn:4pcf_estimator} we notice that, if we seek to remove the BAO feature from the 4PCF, the only relevant term is $\mathcal{N}$, since it is a product of four fields $N$, each of which contain the data field $D(\bfx)$. In contrast, the randoms $R(\bfx)$ do not have BAO imprinted on them. In the next subsection we thus focus on de-wiggling $N$ (and thus $\mathcal{N}$).

\subsubsection{Field-Level Approach and Role of the Green's Function}
As mentioned at the start of this section, to remove the BAO features, the most straightforward method is to operate at the field level: take the Fourier transform of $N(\bfx)$ field, multiply by the ratio of matter to no-wiggle transfer function, inverse Fourier transform the field back in the configuration space, and proceed by computing the 4PCFs as in \cite{Cahn2021:parity} and \cite{hou2022:parity}. While there are more efficient methods available, building a solid understanding of this basic procedure will pave the way for improvements.

In the field-level de-wiggling approach in position space, we treat each late-time galaxy as a proxy for a primordial density perturbation. To understand the evolution of such a perturbation, consider an isolated primordial density perturbation modeled as a Dirac delta distribution. The resulting late-time matter distribution can be interpreted through the matter Green’s function, which depends on both space and time. At late times during the matter-dominated era, the growth rate of perturbations becomes approximately scale-independent and the Green’s function becomes separable in space and time~\footnote{We assume the spatial-temporal separability in the context of Green's function since we are interested in the galaxy clustering at $0<z<2$. However, the separability is not guaranteed {\it e.g.} during reionization, which affects the baryons.}. Dividing the Green’s function by the time-dependent growth factor, the remaining spatial function is equivalent to the inverse Fourier transform of the matter transfer function~\cite{ESW_07}. 
%Figure 1 of~\cite{ESW_07}  shows the behavior of the dark matter and baryon (gas, in their legend) Green's functions; their weighted average can be treated as the matter Green's function, though at late times they converge completely under gravity in any case.  This Figure shows that a primordial density perturbation will produce an extended mass distribution with a BAO feature. We note that the Green's function approach neglects any non-linear gravitational evolution. 

If we want to return the density field around a given point to its primordial state, we can simply apply the time-reversed matter Green’s function around that point. This process can be visualized as taking the sphere of excess galaxies at the BAO scale around a late-time galaxy and evolving it backwards until it collapses to the origin. This effectively converts the late-time matter distribution around a point back to its primordial configuration. Since we are focusing on BAO features, which exist on sufficiently large scales to approximate the effects of gravity as linear, this process can be treated as the time-reversal of a linear differential equation. Consequently, it satisfies the superposition principle, meaning we can apply this transformation independently around every galaxy. In short, we may convolve the time-reversed matter Green’s function with the late-time galaxy density field to approximate the primordial density field~\footnote{The Green's function picture of BAO was first developed in \cite{bash_bert, bashinsky}.}. 

To evolve the primordial field forward as if in a universe without BAO, we can envision baryons as being massless, which remain tightly coupled to the photons. Without mass, baryons do not exert gravitational influence after decoupling, preventing the formation of the BAO feature in the dark matter distribution that typically arises due to baryon gravitational attraction. We then evolve our primordial field forward using this ``massless baryon'' (or ``no wiggle'') transfer function~\footnote{~\cite{SE_simple} presents the ``massless baryon'' (or ``no wiggle'') transfer function and its inverse Fourier transform in Figure 3, lower panel. As shown, its broadband behavior resembles the standard matter Green’s function (including BAO), as seen in Figure 6 of the same reference, but notably lacks any BAO features. Note that the ``Green's function'' in~\cite{SE_simple} refers to the solution of their equation 27, which is the matter Green's function, divided by the growth rate.}. 

In summary, the operation on the late-time galaxy field can be represented as a composition of two Green’s functions: first, we apply the time-reverse of the standard matter Green’s function, and then evolve forward in time using the no-wiggle Green’s function. 

In practice, applying the time reverse of the matter Green’s function is equivalent to deconvolving the field by the Green’s function. By the Convolution Theorem, this operation corresponds to dividing the density field in $k$ space by the Fourier transform of the matter Green’s function. Due to the factorization of spatial and temporal dependence, the spatial component of the Green’s function is the inverse Fourier transform of the matter transfer function. Similarly, convolving with the ``no wiggle'' Green’s function in Fourier space is equivalent to multiplying by the no wiggle transfer function for its spatial component, while the time-dependent components cancel between the numerator and denominator. As a result, we can define a ``de-wiggling'' kernel that captures both  of these operations, as
\begin{align}\label{eqn:def_G}
        G(\bfr) &\equiv \iFT{\calT_{\delta, \rm nw}(k)/\calT_{\delta}(k)}(\bfr) \\
    &\equiv \iFT{\tilde{G}(k)}(\bfr),\nonumber
\end{align}
where the second line defines $\tilde{G}$ as the transfer function ratio. Since this latter depends only on the magnitude $k$, the inverse FT can depend only on $|\bfr|$; in other words, $G(\bfr) = G(|\bfr|)$, and so is isotropic. 
%This property will be of use in what follows. We also note that there is no time dependence, since the matter Green's function run in time-reverse would divide out the linear growth, but the ``no wiggle'' Green's function applied forward would restore it. Put more mathematically, it simply cancels out of the transfer function ratio because of the separation between $D(z)$ and $\calT_{\rm m}(k)$ (or $\calT_{\rm nw}(k)$ assumed in this work (consistent with the focus on just the linear evolution). 

\subsubsection{De-Wiggling at the Estimator Level}
With the de-wiggling  function in hand, we may now compute the de-wiggled field $N^{\rm nw}$ as
\begin{align}\label{eqn:N_convolv}
    N^{\rm nw}({\bf u}) = \int d^3\vec{u}_1\, G( {\bf u} - {\bf u}_1) N({\bf u}_1).
\end{align}

\iffalse
where $\tilde{G}(k)$ is the de-wiggling kernel in Fourier space, which only has scale dependence due to the statistical isotropy.
Consequently, the de-wiggling kernel in the position space has only magnitude dependence
\begin{align}
\label{eqn:G_iso}
    G(\bfr) \equiv G(|\bfr|).
\end{align}
\fi
% The Green’s function describes how initial localized perturbations evolve over time. 
% In the context of BAO, these perturbations evolve into spherical sound waves that propagate outward from the origin of the perturbation. Meanwhile, dark matter converges toward the baryonic matter ({\it e.g.} figure 1 of~\citep{ESW_07}) \footnote{\cite{SE_simple} provides a comprehensive, closed-form analytic description of how BAO features arise using a two-fluid approximation. Their figures 3 and 6 illustrate the density Green’s functions with and without BAO, respectively.}.

We may now project a product of four de-wiggled fields at appropriate points for the 4PCF onto our isotropic basis. We also include radial binning, which enables us to rewrite the angular integrals over the relative positions $\hat{r}_i$ as full 3D integrals. We have
\begin{align}
\label{eqn:dewiggle}
    &\calN^{\rm nw}_{\ell_1 \ell_2 \ell_3}(R_1, R_2, R_3) = \\
    & \int d^3 \vec{x}\; N^{\rm nw} (\vec{x})\prod_{i=1}^3\left[\int d^3 \vec{r}_i\; N^{\rm nw} (\vec{x} + \vec{r}_i) \Theta(r_i; R_i)\right]\nonumber\\
    &\qquad\qquad\qquad\times\mathcal{P}^*_{\ell_1 \ell_2 \ell_3}(\hat{r}_1, \hat{r}_2, \hat{r}_3) \nonumber.
\end{align}
Here $\Theta$ is a binning function ensuring that $r_i$ is within a spherical shell of width $\Delta$ and center $R_i$, and $R_i$ is used to identify the bin. Each bin is normalized to have unit volume.

We now expand the isotropic function in the last line in terms of spherical harmonics using its definition Eq.~\eqref{eqn:Plll}, and replace all the $\bfr_i$-dependent fields $N^{\rm nw}(\bfx+\bfr_i)$ by Eq.~\eqref{eqn:N_convolv}. We obtain
\begin{align}
\label{eqn:dewiggle-2}
    &\calN^{\rm nw}_{\ell_1 \ell_2 \ell_3}(R_1, R_2, R_3) = 
    \int d^3 \vec{x}\;  N^{\rm nw}(\vec{x}) \sum_{m_1 m_2 m_3} (-1)^{\ell_1+\ell_2+\ell_3} \nonumber\\
    &\times \six{\ell_1}{\ell_2}{\ell_3}{m_1}{m_2}{m_3} \prod_{i=1}^3 \Bigg[\int d^3 \vec{r}_i\; \int d^3\vec{u}_i \; G(|\vec{x} + \vec{r}_i - \vec{u}_i|)\nonumber\\  
    & \times\, N(\vec{u}_i)\Theta(r_i; R_i) Y^*_{\ell_i m_i}(\hr_i)\Bigg] \nonumber\\
    & \equiv \int d^3 \vec{x}\;  N^{\rm nw}(\vec{x}) \sum_{m_1 m_2 m_3} (-1)^{\ell_1+\ell_2+\ell_3} \six{\ell_1}{\ell_2}{\ell_3}{m_1}{m_2}{m_3} \nonumber\\
    &\quad\times \prod_{i=1}^3 \calI_{\ell_i m_i}(\bfx, \bfu_i, \bfr_i ; R_i).
\end{align}
We observe that our $N$ fields no longer depend on the $\vec{r}_i$; thus, these integrals may be performed first, independently of the data. We also see that Eq.~(\ref{eqn:dewiggle-2}) contains three factors of the binned, de-wiggled density fields. This motivates introducing
\begin{align}\label{eqn:I_lm}
    &\calI_{\ell_i m_i}(\bfx, \bfu_i, \bfr_i ; R_i) \nonumber\\
    &\equiv \int d^3\vec{r}_i\; G(|(\vec{u}_i-\vec{x})-\vec{r}_i|) \Theta(r_i; R_i) Y^*_{\ell_i m_i}(\hat{r}_i) \nonumber\\
    &=\int d^3\vec{r}_i\, \left[G (\vec{r}_i) \circledast (\Theta Y^*_{\ell_i m_i})(\vec{r}_i)\right] (\vec{u}_i-\vec{x}),
\end{align}
where to make the convolutional structure more evident, in the second line above we rearranged the argument of the de-wiggling function relative to how it appears in Eq.~(\ref{eqn:dewiggle-2} using that $G(|\vec{r}_i + (\vec{x}-\vec{u}_i)|) = G(|(\vec{u}_i-\vec{x})-\vec{r}_i|)$. By using
\begin{eqnarray}
   &&\FT{Y_{\ell_i m_i}(\hr_i)}(\bfk_i)\\ 
   &&\,= 4\pi\, \sum_{\ell_i m_i} (-i)^{\ell_i}\,Y_{\ell_i m_i}(\hk_i)\,\int r_i^2 dr_i\,  j_{\ell_i}(k_i r_i),\nonumber
\end{eqnarray}
the convolution in Eq.~\eqref{eqn:I_lm} can be written as
\begin{align}
    &\left[G (|\vec{r}_i|) \circledast (\Theta Y^*_{\ell_i m_i})(\vec{r}_i) \right](\vec{u}_i-\vec{x})\\
    &= Y_{\ell_i m_i}(\widehat{\vec{u}_i-\vec{x}}) \int \frac{k_i^2 dk_i}{2\pi^2}\, 
    j_{\ell_i}(k_i|\vec{u}_i-\vec{x}|) \tilde{G}(k_i) \tilde{\Theta}_{\ell_i}(k_i; R_i),\nonumber\\
    &\equiv Y_{\ell_i m_i}(\widehat{\vec{u}_i-\vec{x}}) g_{\ell_i}(|\vec{u}_i-\vec{x}|; R_i),
    \nonumber
\end{align}
where the Fourier transform of the binning function is
\begin{eqnarray}
    &\tilde{\Theta}_{\ell_i}(k_i; R_i) \equiv 4\pi \int r^2 dr\, \Theta(r_i; R_i) j_{\ell_i}(k_i r_i).    
\end{eqnarray}
The integral giving $\tilde{\Theta}_{\ell}$ can be performed analytically~\citep{bloomfield}. While there is no analytic solution for $g_{\ell_i}$ because it is a transform of  $\tilde{G}$, which latter involves a ratio of the transfer functions, $g_{\ell_i}$ can be obtained at each $\ell_i$ with a 1D FFTLog \citep{Hamilton, Siegman, Talman, rot_method}.

We may now write 
\begin{align}\label{eqn:dewiggle_2}
&\calN^{\rm nw}_{\ell_1 \ell_2 \ell_3}(R_1, R_2, R_3)\\
&=\int d^3 \vec{x}\; N^{\rm nw}(\vec{x})\sum_{m_1 m_2 m_3} (-1)^{\ell_1+\ell_2+\ell_3} 
\six{\ell_1}{\ell_2}{\ell_3}{m_1}{m_2}{m_3} \nonumber\\
&\times \prod_{i = 1}^3 \int d^3\vec{u}_i\; N(\vec{u}_i) Y_{\ell_i m_i}^*(\widehat{\vec{u}_i -\vec{x}}) g_{\ell_i}(|\vec{u}_i - \vec{x}|)\nonumber\\
&= \int d^3 \vec{x}\; N^{\rm nw}(\vec{x})\sum_{m_1 m_2 m_3}  
\six{\ell_1}{\ell_2}{\ell_3}{m_1}{m_2}{m_3}\nonumber\\
&\,\times\,\prod_{i = 1}^3\bigg[\left[N(\vec{u}_i) \circledast (Y_{\ell_i m_i} g_{\ell_i})(\vec{u}_i)\right](\bfx)\bigg].\nonumber
\end{align}
To obtain the last equality above, we used that the integrals over $\vec{u}_i$ are convolutions as well. Since $Y_{\ell 
 m}(-\hu)=(-1)^{\ell} Y_{\ell m}(\hu)$, the phase $(-1)^{\ell_1+\ell_2+\ell_3}$ ends up cancelling out.

The algorithm for estimating the 4PCF of a field in the isotropic basis is exactly as Eq.~(\ref{eqn:dewiggle_2}) but with $g_{\ell_i}{(|\bfu_i - \bfx|;R_i)} \to \Theta(|\bfu_i - \bfx|; R_i)$, the usual spherical shell binning. Thus, if we treat $g_{\ell_i}(|\bfu_i - \bfx|;R_i)$ as an effective binning function, then Eq.~(\ref{eqn:dewiggle_2}) can be evaluated using the standard NPCF algorithms \citep{SE_3PCF_FT, Portillo, Philcox:encore, sunseri}.

\section{Discussion}
\label{sec:discussion}
Here we further discuss  what might result from application of our de-wiggling procedure in the absence of a true parity-odd 4PCF. We then turn to the possible impact of systematics or other new physics on our approach, and finally, consider use of odd-4PCF BAO as a new standard ruler.
\subsection{Offset in Detection Significance from De-Wiggling  Without Genuine Odd 4PCF}\label{sec:detxn_null}
One concern regarding the proposed de-wiggling procedure is that it leads to a non-vanishing $\Delta \chi^2_{\rm bao}$ even in the absence of a parity-odd 4PCF signal, as already briefly discussed in \S\ref{subsub:bao-sig}. This occurs because de-wiggling alters the spatial configuration of the density field, by convolving the original density field with the de-wiggling function. This change to the density field will then alter the measured 4PCF relative to that of the non-de-wiggled (untouched) field.

%First, if the odd 4PCF arises from systematics rather than underlying physics, the de-wiggling procedure would still be effective, as systematics directly modulate the BAO-imprinted density field, similar to a genuine physical signal. 
We now seek to explore this issue quantitatively. We begin by considering the 4PCF coefficients, given by taking the inverse FT of the trispectrum and projecting onto  the isotropic basis functions:
\begin{eqnarray}
   &&\zeta^{\rm nw}_{\ell_1\ell_2\ell_3}(r_1, r_2, r_3)\nonumber\\
   &=& \int d\hr_1 d\hr_2 d\hr_3 \,\calP^*_{\ell_1\ell_2\ell_3}(\hr_1,\hr_2, \hr_3)\\ 
   &&  \times\, \iFT {(2\pi)^3 \delta_{\rm D}^{[3]}(\bfk_{1234}) \, T(\bfk_1, \bfk_2, \bfk_3) \prod_{i=1}^4  \tilde{G}(k_i) } ,\nonumber
\end{eqnarray}
where $\tilde{G}(k_i)$ is the Fourier-space de-wiggling function (Eq. \ref{eqn:def_G}) and $\bfk_{1234}=\sum_{i=1}^4 \bfk_i$. 

We approximate the ratio of the no-wiggle to the wiggle transfer function as
\begin{eqnarray}
\label{eqn:transfer_ratio}
    \tilde{G}(k) &\approx& 1 - A \left(\frac{k}{k_{\rm bao}}\right) j_0\left(\frac{k}{k_{\rm bao}}\right)e^{-(k/k_{\rm cut})^2}\nonumber\\
    &&\approx 1-\epsilon(k)
\end{eqnarray}
where $A = 0.03$  is a constant amplitude, the  BAO wavenumber $k_{\rm bao}=0.01 \, \hmpc$ and we impose a cut-off scale is $k_{\rm cut}=0.3\,\hmpc$.   As shown in Fig.~\ref{fig:ratio_pkw_pknw_tm_tnw},  this formula gives a give a good empirical fit to the transfer function ratio. If we approximate $\epsilon\approx \calO(1\%)$ as constant across the bins, we may pull $\epsilon$ out of our inverse FT integrals.
% , and $T_{\ell_1\ell_2\ell_3}(k_1, k_2, k_3)$ is the radial coefficient of the full trispectrum in the isotropic basis. 
% We now seek to explore this issue quantitatively. We begin by considering the 4PCF coefficients. These are the multi-dimensional Hankel (\textit{i.e.} triple-sBf) transform of the trispectrum \footnote{This is straightforward to show using the plane wave expansion and orthogonality of the spherical harmonics}:
% \begin{eqnarray}
%    &&\zeta^{\rm nw}_{\ell_1\ell_2\ell_3}(r_1, r_2, r_3) = \\ 
%    && \int \left[\,\prod_{i=1}^3  \frac{k_i^2 dk_i}{2\pi^2}  \tilde{G}(k_i) j_{\ell_i}(k_i r_i) \right] \tilde{G}(k_{123})  T_{\ell_1\ell_2\ell_3}(k_1, k_2, k_3).\nonumber
% \end{eqnarray}
% $k_{123}=|\bfk_1+\bfk_2+\bfk_3|$, $\tilde{G}(k_i)$ is the Fourier-space de-wiggling function (Eq. \ref{eqn:def_G}), and $T_{\ell_1\ell_2\ell_3}(k_1, k_2, k_3)$ is the radial coefficient of the full trispectrum in the isotropic basis. 

% In the absence of genuine parity violation, each coefficient of the odd trispectrum can be regarded as drawn from a multivariate Gaussian distribution, centered at zero and with standard deviation ${\sigma_{\ell_1, \ell_2, \ell_3; k_1, k_2, k_3}}$.
% \begin{eqnarray}
% T_{\ell_1\ell_2\ell_3}(k_1, k_2, k_3) \sim \calN({0}, \sigma_{\ell_1, \ell_2, \ell_3; k_1, k_2, k_3}),  
% \end{eqnarray}
% where $\sigma_i$ is given by the square root of the diagonal of the covariance matrix $\sqrt{\mathbb{C}_{ii})}$, where $i$ denotes the index of the  tetrahedral configuration.

In the absence of genuine parity violation, the set of odd 4PCF coefficients, which we denote with curly brackets, follows a multivariate Gaussian distribution, centered at zero and with covariance matrix $\mathbb{C}$:
\begin{eqnarray}
\{ \zeta_{\ell_1\ell_2\ell_3}(r_1, r_2, r_3) \}\sim \calN({0}, \mathbb{C}).
\end{eqnarray}
\iffalse
%Zack---we don't need to say theaverage over the bin is zero---that already must be true if we are asserting the odd 4PCF coefficients are all centered at zero!
and the average over each bin of the odd 4PCF will be zero
\begin{eqnarray}
    \av{\zeta_{\ell_1\ell_2\ell_3}(r_1, r_2, r_3)}_{\rm bin} = {0},
\end{eqnarray}
where we use $\av{\ldots}_{\rm bin}$ to denote bin-averaging. 
\fi
However, the \textit{variance} of the 4PCF will be changed slightly depending on whether we use with-wiggle or no-wiggle  transfer functions.

We may roughly compute  the variance ratio. The variance of each coefficient is just the expectation value of its square, since the mean of each coefficient is zero, and so we find 
% \begin{eqnarray}\label{eqn:zeta_nw_ratio}
%     && \frac{\text{Var}[\zeta^{\rm nw}]_{\rm bin}}{\text{Var}[\zeta^{\rm w}]_{\rm bin}}\\ 
%     &=& \Bigg(   \frac{\int\left[\prod_{i=1}^3\frac{k_i^2 dk_i}{2\pi^2}  \,\tilde{G}(k_i) \,j_{\ell_i}(k_i r_i) \right]\tilde{G}(k_{123}) T_{\ell_1\ell_2\ell_3}(k_1, k_2, k_3)}{\int \left[\prod_{i=1}^3  \frac{k_i^2 dk_i}{2\pi^2} \,j_{\ell_i}(k_i r_i)T_{\ell_1\ell_2\ell_3}(k_1, k_2, k_3) \right] } \Bigg)^2.\nonumber
% \end{eqnarray}

% Eq.~\eqref{eqn:zeta_nw_ratio} thus simplifies further to
\begin{eqnarray}
    && \frac{\text{Var}[\zeta^{\rm nw}_{\ell_1 \ell_2  \ell_3}(r_1, r_2, r_3)]}{\text{Var}[\zeta^{\rm w}]_{\ell_1 \ell_2  \ell_3}(r_1, r_2, r_3)} 
    \approx (1-4\epsilon)^2,    
\end{eqnarray}
This variance difference propagates into our BAO detection significance  as computed via Eq.~\eqref{eqn:Delta_tilde_chi2_bao}. For a 4PCF data vector with $N$ degrees of freedom, but in the absence of any genuine parity violation, we will find 
\begin{eqnarray}
\label{eqn:delta_chi2_bao_null}
\Delta\tilde{\chi}^2_{\rm bao, null} \approx (4\epsilon)^2 \, N. 
\end{eqnarray}
In this work, we consider $N=960$ degrees of freedom; thus we might expect $\Delta\tilde{\chi}^2_{\rm bao, null}\approx 0.04^2 \times 1,000\!\sim\!\calO(1)$ for the Gaussian, no-parity violation simulations, as indeed we found in Fig.~\ref{fig:stat_dist_4pcf_bao_20x160x10}.
\begin{figure}
    \centering
    \includegraphics[width=.4\textwidth]{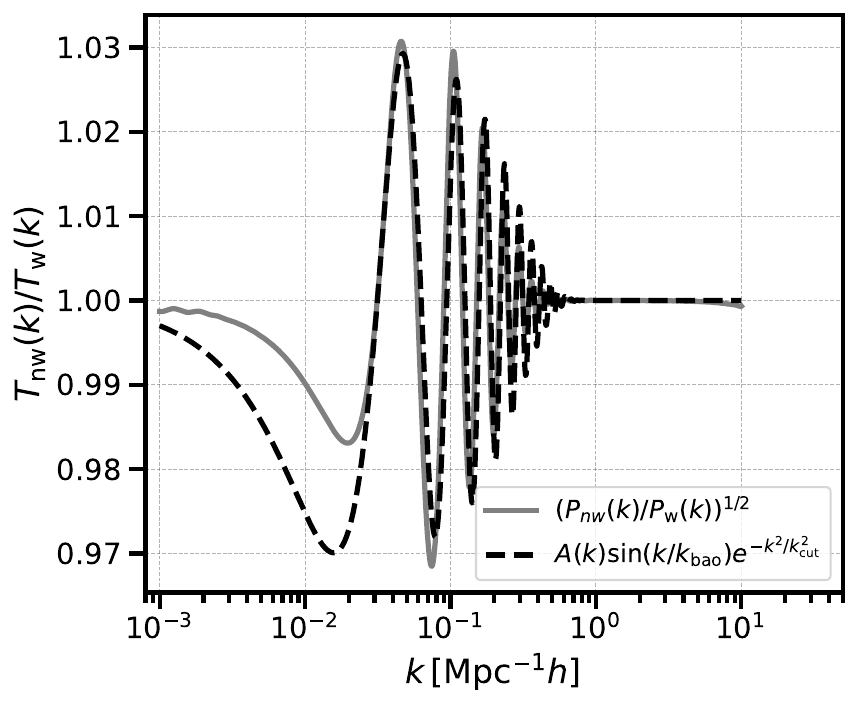}
    \caption{Solid grey: ratio of the transfer function without BAO (``no wiggle'', subscript ``nw'') from \cite{EH_98} to that with BAO (subscript ``w'') from \textsc{CLASS}. Dashed black: our approximation for this ratio, Eq. (\ref{eqn:transfer_ratio}). We see that our approximation describes the ratio rather well.}
    \label{fig:ratio_pkw_pknw_tm_tnw}
\end{figure}
To understand the offset in the detection significance resulting from the de-wiggling procedure, we could alternatively have begun directly with Eq.~\eqref{eqn:Delta_tilde_chi2_bao}, where we now include the possibility of a parity-odd signal. In this scenario, using Eq. \eqref{eqn:transfer_ratio}, the detection significance becomes
% Since the template used in the simulation is smooth at the scales of interest, we can use Eq.\eqref{eqn:transfer_ratio} to further estimate the relation between the detection of the BAO and the overall parity-odd significance. 
\begin{eqnarray}\label{eqn:Delta_chi2_relation}
\av{\Delta\tilde{\chi}^2_{\rm bao}} &\approx&  \av{\left(\zeta^{\rm w}_{\rm d} - (1-4\epsilon)\zeta^{\rm w}_{\rm d}\right) \mathbb{C}^{-1} (\zeta^{\rm w}_{\rm d} - \left(1-4\epsilon)\zeta^{\rm w}_{\rm d}\right)^{\rm T}}\nonumber\\
&=& (4\epsilon)^2 (\Delta{\chi}^2_{\rm odd}+N),
\end{eqnarray}
where we recognize $\Delta\chi^2_{\rm odd}=\zeta^{\rm w}_{\rm m}\mathbb{C}^{-1}\zeta^{\rm w, T}_{\rm m}-N$. For $\Delta\chi^2_{\rm odd}=0$, we recover Eq.~\eqref{eqn:delta_chi2_bao_null}. 

We pause to notice that  Eq.~\eqref{eqn:Delta_chi2_relation} offers a scaling relation between the overall parity-odd detection significance and the BAO significance.
In this work, we found $\Delta{\chi}^2_{\rm odd}\approx 200$ and $\Delta\tilde{\chi}^2_{\rm bao}\approx 1$, which corresponds to $\epsilon\approx \calO(1 \%)$ and is again consistent with the effect of BAO wiggles. This scaling relation could be useful for distinguishing between a signal and under-estimated statistical noise, as the signal is likely scale-dependent, whereas the noise remains constant per degree of freedom and so we would see a linear increase in $\Delta \tilde{\chi}^2_{\rm bao}$ as we add degrees of freedom  (though see footnote~\footnote{We note that the scaling relation does not necessarily allow us to distinguish whether the scale-dependence is due to a genuine signal or systematics. To fully distinguish the two, a parity-violating model is required.}).

\subsection{Systematics and Additional Physics with Possible Impact on Detection Significance}
\label{subsec:sys}
Here, we discuss systematics that could either contaminate or degrade the BAO significance. Given the 
hierarchical nature of large-scale structure (LSS) correlation functions, systematics that induce a spurious sharp feature at scales in the 4PCF would likely already have a corresponding effect in the 2PCF and 3PCF, and thus be revealed (and removable using)  these lower-order statistics. However, the 4PCF likely has a lower signal-to-noise ratio than 2PCF and 3PCF, so could be more adversely affected by issues that would not be noticed in these latter.

We now  briefly outline several possible issues to consider when performing the odd-4PCF BAO search.
\begin{itemize}
    \item{Non-linear evolution and redshift-space distortions (RSD)}: gravitational evolution introduces non-linearity in the matter distribution, and the de-wiggling method outlined in \S\ref{sec:dewiggling} does not address this. However, it is important to note that such non-linear effects are important primarily on scales well below the BAO scale, and in studies of the 2PCF have not been found to greatly influence either the BAO detection significance or its use as a standard ruler(for instance \cite{ESW_07} Section 3). Furthermore, to undo the non-linear gravitational evolution, density field reconstruction could be added as a pre-processing step \cite{Recon_07,Recon_09,White2015:ReconZD,Burden2015:ReconFFT,Schmittfull2017:InterBAORecon,Birkin2019:ReconBAO}. Doing so would offer a more linear field before de-wiggling, sharpen the BAO feature, and remove large-scale RSD.
\item{Baryon-dark matter relative velocity bias}: One of the few known physical effects that can modulate galaxy clustering on BAO scales is the relative velocity between baryons and dark matter, due to the fact that they have different behaviors prior to decoupling~\citep{tsel}. Though decaying after $z\sim 1,020$, this relative velocity could impact the formation of the high-redshift early galaxies that are the ancestors of the galaxies used for late-time spectroscopic  samples; this effect is captured by adding a term to the galaxy bias expansion ~\citep{yoo, se_rv}. Thus, the BAO feature can be shifted in position and have its shape slightly altered \citep{dalal_10}. Extensive work has been carried out using the BOSS with the 3PCF \citep{se_rv, se_rv_boss} and the power spectrum \citep{beut_rv}. These works did not find a non-zero relative velocity bias in BOSS, and were able to constrain it sufficiently that any effect on the BAO scale measured from BOSS is  far less than the statistical error-bars. 
\item{Neutrinos}: Cosmic neutrinos have two impacts on LSS. First, they cause a suppression of the overall clustering amplitude, which is a smooth, broadband effect and does not lead to any  sharp features at the BAO scale. Second, prior to the formation of the CMB, the neutrinos in the early Universe are relativistic; they thus alter the phase of the BAO oscillations~\citep{bashinsky, baumann-phases}. In position space, this slightly changes the shape of the BAO feature (shown \cite{baumann-phase-detxn} Figure 6, lower panel); this is  a very small effect, and could  certainly be folded  into future odd-4PCF BAO analyses if desired.

\item{Primordial features}: Various primordial features induced by beyond-single-field slow-roll inflationary models could in principle impact the BAO. In particular, isocurvature perturbations, which are fluctuations in the relative number densities of different species (but holding the overall density unchanged), have been considered in this context~\citep{Linde1997:IsoCP}. While large isocurvature perturbations have been ruled out by the CMB ~\citep{Hikage2009:IsoCPWMAP,Planck18:CosmologyPNG}, relatively significant baryon fluctuations could still exist as long as they are compensated by the dark matter fluctuations. This latter scenario is called compensated isocurvature perturbations (CIP), and could modulate the BAO scale from region to region due to local changes in the sound speed (which depends on the local baryon density) ~\citep{Heinrich2019:IsoCPBAO}. However, CMB measurements are insensitive to CIP, and as yet no detection of CIP has been reported from galaxy surveys~\citep{Barreira2023:CIPBOSSDR12}. Certainly this effect could be  constrained from standard even-sector BAO searches with 2- and 3-point statistics if it were to become a concern for the odd-4PCF BAO search.

\item A late-time parity-violating mechanism: While all primordial parity-violating mechanisms would ensure BAO imprints on the odd sector, this is not necessarily true for late-time mechanisms. For instance, one could imagine a late-time parity-violating mechanism that produces density perturbations that are independent from the normal, adiabatic perturbations. These new perturbations would then not contain BAO, but via gravity, they could still impact the galaxy distribution and create an parity-odd 4PCF. Ultimately this possibility is as yet speculative and requires further theoretical development.

\iffalse
However, we note that the de-wiggling procedure in our paper would still create a non-vanishing $\Delta\tilde{\chi}^2$, only with a small correction $128\epsilon^3 \Delta\chi^2_{\rm odd}$\footnote{The estimation is identical to Eq.~\eqref{eqn:Delta_chi2_relation}, except for replacing $\zeta_{\rm m}^{\rm w} \rightarrow \zeta_{\rm m}^{\rm nw} \approx (1-4\epsilon)\zeta_{\rm m}^{\rm w}$.}. To address this issue, a specific parity-violating model would be necessary to aid the analysis.
\fi
\end{itemize}

%cite ashley ao systematics paper \citep{ross_syst_17, merz, ding, eisen_theo_bao}

\subsection{BAO from the Parity-Odd Sector as a Standard Ruler}

The BAO have been used as a standard ruler to infer cosmological distances~\citep{EHT_98, Blake_03, HH_03, Linder_03, ESW_07, SE_3PCF_BAO, lado_ho}. The precision of these distance measurements is typically quantified by the error-bar on a dilation parameter, $\alpha$, which relates the BAO scale observed from a galaxy survey to that predicted in a fiducial cosmological model. If the observed BAO scale matches that predicted, $\alpha = 1$. Using this idea, if BAO are detected in the parity-odd sector, they can be used as a standard ruler.

We now examine the expected error-bar $\sigma(\alpha)$ on the dilation parameter. We begin with the Fisher matrix, with elements
\begin{eqnarray}
\label{eqn:fisher}
    F_{\mu \nu}=\sum_{i, j=1}^{N} \frac{\partial T_i}{\partial \theta_\mu} \mathbb{C}_{i j}^{-1} \frac{\partial T_j}{\partial \theta_\nu},
\end{eqnarray}
where $i$ and $j$ denote trispectrum configurations (\textit{i.e.} $\bfk_1, \bfk_2,\bfk_3$), $N$ is the number of degrees of freedom, and $\theta_{\mu}$ and $\theta_{\nu}$ are elements of a vector $\theta$ made up of the parameters under consideration. 

Here  we consider only 
$\alpha$ so the parameter vector has length one in Eq. (\ref{eqn:fisher}), and the matrix $F_{\mu \nu}$ has just one element, which we denote $F$. The error on $\alpha$ is then
\begin{eqnarray}
\label{eqn:error_fisher}
\sigma(\alpha)=1/\sqrt{F}.  
\end{eqnarray}
To obtain $F$ we require $\partial T_i/\partial \alpha$. One can easily show that, for the $i^{\rm th}$ trispectrum configuration, 
\begin{align}
    \left[ \frac{\partial \, \ln \, T}{\partial \alpha }\bigg|_{\alpha = 1} \right]_i = \sum_{p = 1}^4 \left[  \frac{\partial\, \ln\,\calT_\delta(k_p(\alpha))}{\partial \alpha}\bigg|_{\alpha = 1}\right]_i
\end{align}
where $k_p \equiv |\bfk_p|$ and the notation $[ \cdots ]_i$  means that these four wavenumber magnitudes correspond to the $i^{th}$ configuration. The transfer function is a function of the ``apparent'' wavenumber, assuming a fiducial cosmology. The ``apparent'' wavenumber, in turn, depends on the dilation parameter $\alpha$, with $k(\alpha)=\alpha k'$ denoting the scaling of the wavenumber when applying the dilation parameter to the fiducial grid. Multiplying both sides of the above relation by $T_i$ (evaluated at $\alpha = 1$) then gives the desired $\partial T_i/\partial \alpha$.

We then have 
\begin{align}
\label{eqn:fisher-F}
F &= \sum_{i, j =1}^N  T_i \mathbb{C}_{ij}^{-1} T_j 
\bigg\{\sum_{p, q = 1}^4 
\left[  \frac{\partial\, \ln\,\calT_\delta(k_p(\alpha))}{\partial \alpha}\bigg|_{\alpha = 1}\right]_i\nonumber\\
&\qquad \qquad \qquad\qquad\,\times\left[  \frac{\partial\, \ln\,\calT_\delta(k_p(\alpha))}{\partial \alpha}\bigg|_{\alpha = 1}\right]_j\bigg\}.
\end{align}
We notice that the first factor in the outer sum  above, $T_i \mathbb{C}_{ij}^{-1} T_j$, in the limit that the quantity in curly brackets went to a constant, would simply lead to $F$ proportional to the \textit{overall} odd-trispectrum detection significance

In detail, in this limit $F$ would just become $\chi^2$ for the trispectrum (\textit{cf.} Eq. \ref{eqn:chi-squared-twelve}) with the model set to zero, as corresponds to our null case of no parity violation. The detection significance is $S_{\rm odd} \approx \chi^2/\sqrt{2N}$, since the $\chi^2$ distribution with $N$ degrees of freedom  has width $\sqrt{2N}$ \cite{Cahn2021:parity}. 
 Hence  in this limit $F \propto S_{\rm odd}$. We will return to this point.

We also note that
\begin{align}
    \frac{\partial\, \ln\,\calT_\delta(k_p(\alpha))}{\partial \alpha}\bigg|_{\alpha = 1} = 
    \frac{\partial\, \ln\,\calT_\delta(k)}{\partial k}\bigg|_{k = k_p} k_p.
    \label{eqn:chain-rule}
\end{align}
The transfer function may  be written  \cite{EH_98} (their Eq. 16) as 
\begin{align}
    \calT_{\delta}(k) = (1 - f_{\rm b}) \calT_{\rm c}(k) + f_{\rm b}\calT_{\rm b}(k),
    \label{eq:eh98-motiv-form}
\end{align}
where $f_{\rm b} \approx 0.2$ is the baryon fraction (with respect to the total matter), $\calT_{\rm c}$ is the Cold Dark Matter (CDM) transfer  function, and $\calT_{\rm b}$ is the baryon transfer function. A very rough form for $\calT_{\rm c}$, motivated by \cite{EH_98}, is
\begin{align}
    \calT_{\rm c}(k) \simeq \frac{1}{1 + (k/k_{\rm eq})^2},
\end{align}
where $k_{\rm eq} \approx 0.01 \hmpc$ is the scale of the horizon at matter-radiation equality. Meanwhile, it is also the wavenumber at which the BAO wiggles in the transfer function begin, and then continue towards larger $k$.

We then have, dropping the baryon term in Eq. (\ref{eq:eh98-motiv-form})  entirely (since $f_{\rm b} = 0.2$), that
\begin{align}
    \frac{\partial\, \ln\,\calT_\delta(k)}{\partial k} \approx
    \frac{\partial\, \ln\,\calT_{\rm c} (k)}{\partial k} \simeq 
     -\frac{2k/k_{\rm eq}^2}{1 + (k/k_{\rm eq})^2}  \to  -\frac{2}{k},
\end{align}
where the final, limiting case is for $k \gg k_{\rm eq}$. 

Inserting this limiting form in Eq. (\ref{eqn:chain-rule}) and evaluating it at $k_p$, we see that the $k_p$ dependence cancels out and we find
\begin{align}
    \frac{\partial\, \ln\,\calT_\delta(k_p(\alpha))}{\partial \alpha}\bigg|_{\alpha = 1} &= 
    \frac{\partial\, \ln\,\calT_\delta(k)}{\partial k}\bigg|_{k = k_p} k_p \nonumber\\
    &\approx -\frac{2}{k}\bigg|_{k = k_p}  k_p = -2.
    \label{eqn:const-transf}
\end{align}
Interestingly, since the BAO wiggles in the transfer function begin at $k \approx k_{\rm eq}$ and continue to higher $k$, and this is also where the transfer function begins to have a non-trivial shape  (for $k \ll k_{\rm eq}$ it is rather flat), this approximation may not be unreasonable to very roughly estimate the precision on $\alpha$. 

Using Eq. (\ref{eqn:const-transf}) in Eq. (\ref{eqn:fisher-F}), the whole term in curly brackets there will now be a constant, independent of wavenumber, and so we find that
\begin{align}
    F \propto \sum_{i, j  = 1}^N T_i \mathbb{C}_{ij}^{-1} T_j \propto S_{\rm odd}, 
\end{align}
recalling our discussion below Eq. (\ref{eqn:fisher-F}).
But this simply says that $F$ is proportional to our \textit{overall} parity-odd detection significance. Hence, very roughly we might expect that the precision on $\alpha$, from Eq. (\ref{eqn:error_fisher}), is
\begin{align}
    \sigma(\alpha) = 1/\sqrt{F} \propto S_{\rm odd}^{-1/2};
\end{align}
\textit{i.e.} that the error-bar on $\alpha$ will decrease as the square-root of our overall detection significance grows. 

This formula should be taken simply as a very rough guide; we now offer two caveats. First, we neglected entirely the baryonic piece of the transfer function, arguing  that $f_{\rm b} \ll 1$; however of course the wiggles in this piece carry significant additional BAO information. From this perspective, our rough formula here is likely conservative and one might  in fact hope to do better on $\sigma(\alpha)$. Second, we did not consider any other parameters, such as galaxy bias or $H_0$, that could be degenerate with an attempt to constrain $\alpha$ using broadband shape information from the transfer function. Further exploration of these issues is left for future work.

Finally, we highlight that our work here holds for both the parity-even and parity-odd sectors. Further, in the standard cosmological paradigm, there should be  no cross-covariance between the parity-even and odd sectors. The parity-odd 4PCF can thus add independent information to parameter inference that exploits BAO as a standard ruler.

%Moreover, the uncertainty in BAO detection is inversely related to the overall signal: stronger significance reduces uncertainty, as expected, and vice versa.

\section{Summary}\label{sec:summary}
In this work, we have demonstrated that BAO can imprint on the parity-odd 4PCF. We quantified the significance of both the overall parity-odd 4PCF and the BAO in a parity-violating toy model. We also found that the choice of method to obtain the covariance matrix has a negligible impact. 

We presented a procedure for efficient BAO de-wiggling at the level of the 4PCF itself, and discussed subtleties regarding its application in a Universe without genuine parity violation. We also outlined potential systematics that should be considered when performing the odd-parity BAO search on real data. 

We then pointed out that BAO in the odd sector can be used as a standard ruler to infer cosmological distances, finding a simple rough formula relating the precision on the BAO scale to the overall odd-parity detection significance.

We now conclude with a few final, important points.

(1) While this paper focused on an inflationary origin for the possible parity violation, the approach to significance quantification we present would also apply to post-inflationary parity-violating mechanisms, provided these mechanisms involve some transformations of the linear matter density field. As long as they do, since that density field has BAO, any odd 4PCF arising from transformation of it will as well unless the specific mechanism somehow conspires to remove the BAO feature.

(2) As we have seen, the BAO significance depends on the overall detection significance of the odd signal. Therefore, simultaneous detection of both an overall parity-odd 4PCF, and BAO in it, does not necessarily guarantee
that the systematics or covariance matrix have been perfectly calibrated. In particular, under-estimate of the covariance matrix could  lead to a spurious overall detection along with a spurious BAO detection.

(3) We explored the possible significance of BAO using a suite of parity-violating simulations. In the real Universe, however, the significance will depend on the behavior of the true underlying model with trispectrum configurations, as well as its overall amplitude. Our analysis also assumes linear gravitational evolution and linear galaxy biasing, while the true galaxy bias scheme is likely more complicated ((\textit{e.g.} \cite{desjacques-rev}). Finally, we do not account for the discrete nature of galaxies, which leads to Poisson noise in the observed 4PCF. 

Future work will further explore the details needed to put our method into practice, including addressing these challenges. With development, the method of this work will be well-suited for application to data from SDSS BOSS and extended BOSS (eBOSS), Dark Energy Spectroscopic Instrument~\citep{DESI2016:PartI}, Euclid~\citep{Euclid2011:WhitePaper}, and future missions such as Spherex~\cite{SPHEREx2018} and Roman~\citep{Wang2022:RomanHLSS}.

\section*{Acknowledgments}
We thank Robert N. Cahn and Eiichiro Komatsu for useful discussions. JH thanks~{\hypersetup{urlcolor=black}\href{https://github.com/Moctobers/Acknowledgement/blob/main/fox_in_office.jpg}{Jue Fox}} for his office support. ZS especially thanks Daniel Eisenstein for useful conversations, as well as Elizabeth Lada and Sylvain Mouton. JH has received funding from the European Union’s Horizon 2020 research and innovation program under the Marie Sk\l{}odowska-Curie grant agreement No. 101025187.

\appendix

\section{Modified Statistic for BAO  Significance $\Delta \tilde{\chi}^2$}
\label{appendix:BAO_detxn}
The usual approach to computing the BAO detection significance is to take the $\chi^2$ difference between that for a model without BAO wiggles (in Fourier space), denoted ``nw'', and one with, denoted ``w'' (\textit{e.g.} \cite{Eisenstein_05,Cole_05,SE_3PCF_BAO,Pearson_18}).  We have
\begin{align}
\label{eqn:delta_chi2}
    &\Delta \chi^2_{\rm bao} = \chi^2_{\rm nw}(\zeta_{\rm m}^{\rm nw}) - \chi^2_{\rm w}(\zeta_{\rm m}^{\rm w}),
\end{align}
where $\chi^2_{\rm nw}$ and $\chi^2_{\rm w}$ depend, respectively, on the 4PCF model with wiggles, $\zeta_{\rm m}^{\rm w}$, and the 4PCF model  without wiggles, $\zeta_{\rm m}^{\rm nw}$. This approach has been used in 2-point and 3-point statistics BAO searches (\textit{e.g.} \cite{Eisenstein_05, Cole_05, SE_3PCF_BAO, Pearson_18}). It is justified  by assuming that the data vector, here, the observed, binned correlation function, follows a multi-variate Gaussian. This latter assumption is justified because as long as the bins are not very narrow, the Central Limit Theorem may be invoked on each bin. For the 4-point statistics, we retain this justification, so the binned 4PCF coefficients should follow a multi-variate Gaussian and hence a $\chi^2$ test remains  appropriate.

% For multivariate Gaussian variables, we can expand the likelihood around  $\zeta_0$ up to the second-order
% \begin{eqnarray}
%     \ln\calL(\zeta(\alpha)) = \ln\calL(\zeta_{\rm nw}(\alpha)) + \frac{1}{2} \sum_{ij} \Delta\zeta_i\left.\frac{\partial^2 \ln {\calL}}{\partial \zeta_i \partial \zeta_j}\right|_{\zeta_{\rm nw}} \Delta\zeta_j,
% \end{eqnarray}
% where $\Delta\zeta\equiv \zeta-\zeta_0$ denotes the difference between the two data vectors evaluated at the same dilation parameter. The error bar of a given parameter is associated with the Fisher information matrix given in Eq.~\eqref{eqn:error_fisher}.
% Considering a single parameter, the difference in the log-likelihood is 
% \begin{eqnarray}
%     -2\Delta \ln\calL(\theta)=\Delta\chi^2=\frac{\Delta\theta^2}{ \sigma_\theta^2} \equiv S^2.
% \end{eqnarray}
% In the context of BAO search, $\Delta\chi^2=\Delta\chi^2_{\rm bao}$ and $S=S_{\rm bao}$. Therefore, $\sqrt{\Delta\chi^2_{\rm bao}}$ represents the detection significance.
\subsubsection{Expectation Values}
We first compute $\left<\Delta \chi^2_{\rm bao}\right>$ for the standard BAO search approach. We rewrite the observed 4PCF of the data, $\zeta_{\rm d}$, as the sum of the with-wiggle model $\zeta_{\rm m}^{\rm w}$ (since in the standard approach, we assume that the BAO are real, so the data vector is a noisy realization of a with-wiggle model) and the statistical noise $\varepsilon$:
\begin{align}
\zeta_{\rm d} = \zeta_{\rm m}^{\rm w} + \varepsilon.
\end{align}
We then find
\begin{align}
\label{eqn:chi2_bao_standard}
    &\av{\Delta\chi^2_{\rm bao}} \\
    &= \av{(\zeta_{\rm d}-\zeta^{\rm nw}_{\rm m}) \mathbb{C}^{-1} (\zeta_{\rm d}-\zeta^{\rm nw}_{\rm m})^{\rm T}-(\zeta_{\rm d}-\zeta^{\rm w}_{\rm m}) \mathbb{C}^{-1} (\zeta_{\rm d}-\zeta^{\rm w}_{\rm m})^{\rm T}}\nonumber\\ 
    &= \av{(\Delta\zeta_{\rm m}+\varepsilon) \mathbb{C}^{-1} (\Delta\zeta_{\rm m}+\varepsilon)^{\rm T} - \varepsilon\, \mathbb{C}^{-1} \varepsilon^{\rm T}} \nonumber\\
    &= \Delta\zeta_{\rm m}\,\mathbb{C}^{-1} \Delta\zeta_{\rm m}^{\rm T},\nonumber
\end{align}
where in the last line above we defined  $\Delta\zeta_{\rm m} \equiv \zeta_{\rm m}^{\rm w} - \zeta_{\rm m}^{\rm nw}$ and dropped the expectation value brackets because the wiggle and no wiggle models are deterministic.  

We now make a remark on the reason that, in the standard approach, the detection significance $S_{\rm bao}$ is simply taken as $\sqrt{\Delta \chi^2_{\rm bao}}$. If the ``signal'' is defined as the difference between the wiggle and no-wiggle models, \textit{i.e.} $\Delta\zeta_{\rm m}$, then Eq.~\eqref{eqn:chi2_bao_standard} is simply the square of the associated signal-to-noise ratio. The signal-to-noise ratio is then simply $\sqrt{\Delta\chi^2_{\rm bao}}$, and indicates the number of standard deviations $\sigma$ at which the signal is detected. 

We now turn to computing the expectation value of our modified search statistic, defined in Eq.~\eqref{eqn:Delta_tilde_chi2_bao}. Both terms in the first factor in Eq.~\eqref{eqn:Delta_tilde_chi2_bao} will contain noise:
\begin{align}
    \Delta\tilde{\chi}^2_{\rm bao}
    = \left([\zeta_{\rm  m}^{\rm nw} + \varepsilon^{\rm nw}]-[\zeta_{\rm  m}^{\rm w} + \varepsilon^{\rm w}] \right)\mathbb{C}^{-1}(\cdots)^{\rm T},
\end{align}
where $(\cdots)$ indicates the same as in the first parentheses. In addition to our previous relation for $\zeta_{\rm d}$, where we now must specify that the noise is that appropriate to a ``with wiggle'' universe, we noticed that our de-wiggling procedure will alter the statistical fluctuations in the observed 4PCF as well:
\begin{align}
    \zeta_{\rm d}^{\rm nw} = \zeta_{\rm m}^{\rm nw} + \varepsilon^{\rm nw}.
\end{align}
We then have 
\begin{align}
\label{eqn:delta_chi2_2}
    &\av{\Delta\tilde{\chi}^2_{\rm bao}}\nonumber\\ 
    &= \av{(\zeta_{\rm m}^{\rm nw} - \zeta_{\rm m}^{\rm w} + \varepsilon^{\rm nw} - \varepsilon^{\rm w})\,\mathbb{C}^{-1} (\zeta_{\rm m}^{\rm nw} - \zeta_{\rm m}^{\rm w} + \varepsilon^{\rm nw} - \varepsilon^{\rm w})^{\rm T}}\nonumber\\
    &\equiv \av{(\Delta\zeta_{\rm m} + \Delta\varepsilon)\,\mathbb{C}^{-1} (\Delta\zeta_{\rm m} + \Delta\varepsilon)^{\rm T}}\nonumber\\
    &\approx \av{\Delta\zeta_{\rm m}\,\mathbb{C}^{-1} \Delta\zeta_{\rm m}^{\rm T}} + 16 \left<\epsilon\right>^2 N.
\end{align}
To obtain the second term in the last line above, we needed to evaluate $\Delta \varepsilon \mathbb{C}^{-1} \Delta \varepsilon^{\rm T}$. To do so, we notice that $\varepsilon$ is the noise on the 4PCF, and this will involve four transfer functions. Thus, the difference in  noises between wiggle and no-wiggle will involve a difference of two quantities that each contain a product of four transfer functions, either with or without wiggles. We  may then Taylor-expand about the with-wiggle case (as appropriate  since this  is the  case with which the covariance  will be evaluated in the search), using our Eq. (\ref{eqn:transfer_ratio}). This  leads to $\Delta \varepsilon \approx 4 \left<\epsilon \right> \varepsilon^{\rm w}$,  where here we replace the $k-$dependent factor $\epsilon(k)$ of Eq. (\ref{eqn:transfer_ratio}) with a rough average over all scales, which will  be $\mathcal{O} (1\%)$.  We  then have that
\begin{align}
    \left<\Delta \varepsilon \mathbb{C}^{-1}  \Delta \varepsilon^{\rm T} \right> \approx  16 \left<\epsilon \right>^2 N
\end{align}
as used in the last line of Eq. (\ref{eqn:delta_chi2_2}). 
Hence, comparing Eqs. (\ref{eqn:chi2_bao_standard}) and (\ref{eqn:delta_chi2_2}), we see that
\begin{eqnarray}
    \av{\Delta{\tilde{\chi}}^2_{\rm bao}} \approx \av{\Delta{\chi}^2_{\rm bao}} + 16 \left<\epsilon \right>^2 N.
\end{eqnarray}
We may rewrite this as
\begin{align}
    \av{\Delta{\tilde{\chi}}^2_{\rm bao}} \approx  S_{\rm bao, true}^2 \left[1 + \frac{16 N \left<\epsilon\right>^2}{S_{\rm bao, true}^2}  \right],
\end{align}
where ``true'' indicates the true BAO detection significance if we could perform a standard search. Now, in  the regime of interest, where we actually are able to detect BAO at say $\gtrsim 5\sigma$, the difference in $\Delta \chi^2$ between our method and the standard approach will be negligible as long  as $N \ll 16,000$ degrees of freedom. 
  
For analyses with more degrees of freedom, or in the low significance limit, the offset will be more significant, so in those scenarios we suggest quantifying the significance by comparing with simulations with no parity violation, as discussed, and employed in our toy-model study, in the main text (\S\ref{subsub:bao-sig}).

\subsubsection{Variances}
We can also compute the variance of our modified BAO search statistic and compare it to the variance of the usual $\Delta \chi^2_{\rm bao}$. To compute the variance, we start by calculating the mean of the square of $\Delta\chi^2_{\rm bao}$:
\begin{eqnarray}
    \av{\left(\Delta\chi^2_{\rm bao}\right)^2} &=& \av{\left((\Delta\zeta_{\rm m}+\varepsilon) \mathbb{C}^{-1} (\Delta\zeta_{\rm m}+\varepsilon)^{\rm T} - \varepsilon\, \mathbb{C}^{-1} \varepsilon^{\rm T}\right)^2}\nonumber\\
    &=& \av{\left((\Delta\zeta_{\rm m}+\varepsilon) \mathbb{C}^{-1} (\Delta\zeta_{\rm m}+\varepsilon)^{\rm T}\right)^2} +\av{\left(\varepsilon\, \mathbb{C}^{-1} \varepsilon^{\rm T}\right)^2}\nonumber\\
    &&- \av{(\Delta\zeta_{\rm m}+\varepsilon) \mathbb{C}^{-1} (\Delta\zeta_{\rm m}+\varepsilon)^{\rm T} \varepsilon\, \mathbb{C}^{-1} \varepsilon^{\rm T}}\nonumber\\
    &&- \av{\varepsilon\, \mathbb{C}^{-1} \varepsilon^{\rm T}(\Delta\zeta_{\rm m}+\varepsilon) \mathbb{C}^{-1} (\Delta\zeta_{\rm m}+\varepsilon)^{\rm T} }
\end{eqnarray}
The odd moments of the noise-dependent terms vanish, and we apply Isserlis' (Wick's) theorem to get the even moments. The expression above then becomes
\begin{eqnarray}\label{eqn:delta_chi2_bao_sq}
    \av{\left(\Delta\chi^2_{\rm bao}\right)^2} 
    &=& \left( \Delta\zeta_{\rm m} \mathbb{C}^{-1} \Delta\zeta_{\rm m}^{\rm T}  \right)^2 + 6 N^2\left( \Delta\zeta_{\rm m} \mathbb{C}^{-1} \Delta\zeta_{\rm m}^{\rm T}  \right) + N(N+2)\nonumber\\
    &&\,-2 (\Delta\zeta_{\rm m}^2 N^2 + N(N+2) ) +N(N+2)\nonumber\\
    &=& \left( \Delta\zeta_{\rm m} \mathbb{C}^{-1} \Delta\zeta_{\rm m} ^{\rm T}  \right)^2 + 4 N^2 \left( \Delta\zeta_{\rm m} \mathbb{C}^{-1} \Delta\zeta_{\rm m} ^{\rm T}  \right),
\end{eqnarray}
where we have used the fact that
\begin{eqnarray}
    \av{\left(\varepsilon\, \mathbb{C}^{-1} \varepsilon ^{\rm T}\right)^2} &=&  \text{Var}\left(\varepsilon\, \mathbb{C}^{-1} \varepsilon ^{\rm T}\right) + \av{\varepsilon\, \mathbb{C}^{-1} \varepsilon ^{\rm T}}^2 \nonumber\\
    &=& N(N+2).  
\end{eqnarray}
Here the mean of $\chi^2$ is $\av{\varepsilon\, \mathbb{C}^{-1} \varepsilon ^{\rm T}}= N$ and variance of a $\chi^2$ distribution is $\text{Var}\left(\varepsilon\, \mathbb{C}^{-1} \varepsilon ^{\rm T}\right)=2N$. The variance of $\Delta{\chi}^2_{\rm bao}$ is then
\begin{eqnarray}
    \text{Var}\left(\Delta{\chi}^2_{\rm bao}\right) = 4 N^2 \left( \Delta\zeta_{\rm m} \mathbb{C}^{-1} \Delta\zeta_{\rm m} ^{\rm T}  \right)
\end{eqnarray}

For our modified statistic, the square of the mean of $\Delta\tilde{\chi}^2_{\rm bao}$ is
\begin{align}\label{eqn:delta_tilde_chi2_bao_sq}
    \av{\left(\Delta\tilde{\chi}^2_{\rm bao}\right)^2} 
    &= \av{\left((\Delta\zeta_{\rm m} + \Delta\varepsilon)\,\mathbb{C}^{-1} (\Delta\zeta_{\rm m} + \Delta\varepsilon)^{\rm T}\right)^2}\nonumber\\
    &= \av{\left(\Delta\zeta_{\rm m}\mathbb{C}^{-1} \Delta\zeta_{\rm m}^{\rm T}\right)^2} + \av{\left(\Delta\varepsilon\mathbb{C}^{-1} \Delta\varepsilon^{\rm T}\right)^2} \nonumber\\
    &\quad+ 2\av{\Delta\zeta_{\rm m}\mathbb{C}^{-1} \Delta\zeta_{\rm m}^{\rm T} \,\Delta\varepsilon\mathbb{C}^{-1} \Delta\varepsilon^{\rm T}} \nonumber\\
    &\quad+ 2\av{\Delta\varepsilon\mathbb{C}^{-1} \Delta\zeta_{\rm m}^{\rm T} \,\Delta\zeta_{\rm m}\mathbb{C}^{-1} \Delta\varepsilon^{\rm T}}\nonumber\\
    &\quad+ 2\av{\Delta\zeta_{\rm m}\mathbb{C}^{-1} \Delta\varepsilon^{\rm T} \,\Delta\varepsilon\mathbb{C}^{-1} \Delta\zeta_{\rm m}^{\rm T}}\nonumber\\
    &={\left(\Delta\zeta_{\rm m}\mathbb{C}^{-1} \Delta\zeta_{\rm m}^{\rm T}\right)^2} + 16\epsilon^4 N(N+2) \nonumber\\
    &\quad+ 96\epsilon^2 N\av{\Delta\zeta_{\rm m}\mathbb{C}^{-1} \Delta\zeta_{\rm m}^{\rm T}}, 
\end{align}
where we again used Isserlis' (Wick's) theorem to simplify the mixed terms that involve both $\Delta\zeta_{\rm m}$ and $\Delta\varepsilon$. The variance of $\Delta\tilde{\chi}^2_{\rm bao}$ is then
\begin{eqnarray}
    \text{Var}\left(\Delta\tilde{\chi}^2_{\rm bao}\right) &=& 64\epsilon^2 N\av{\Delta\zeta_{\rm m}\mathbb{C}^{-1} \Delta\zeta_{\rm m}^{\rm T}}\nonumber\\
    &&\,+ 16\epsilon^4 N(N+2)
\end{eqnarray}

% Subtracting the square of the mean in Eq.~\eqref{eqn:chi2_bao_standard} and Eq.~\eqref{eqn:delta_chi2_2} from Eq.~\eqref{eqn:delta_chi2_bao_sq} and   Eq.~\eqref{eqn:delta_tilde_chi2_bao_sq}, in the limit that $64\epsilon^2 N \ll 1$, we find
% \begin{eqnarray}
% \text{Var}(\Delta\tilde{\chi}^2_{\rm bao}) \approx   \text{Var}(\Delta\chi^2_{\rm bao}).
% \end{eqnarray}

\section{Impact of Covariance Matrix}
\label{appendix:cov}
Here we show the impact of the choice of covariance matrix on the detection significance. We consider two covariance matrices: the first is from the analytic template based on assuming the density field is Gaussian Random~\citep{Hou2022:AnalytCov}, while the second is empirically computed from 5,000 Gaussian simulations.

In this work, we do not account for late-time effects such as non-linear gravitational evolution, non-linear galaxy biasing, and redshift-space distortions. Therefore, both the analytic template and the Gaussian simulation-estimated covariances should be suitable. 

We note that our coupling constant is set to be $g=\pm 2 \times 10^7$ so that it is small enough not to render the likelihood non-Gaussian.  We also avoid using overly fine radial bins, setting $\Delta r> 5\,\mpch$, so that they are not so narrow as to render the Central Limit Theorem inapplicable when we use it to argue that the coefficient on each bin is Gaussian-distributed.

Fig.~\ref{fig:stat_dist_4pcf_bao_20x160x10} shows the distribution of the $\Delta\tilde{\chi}^2_{\rm bao}$, with analytic covariance matrix on the left panel and the mock covariance on the right panel. In both cases, we show the positive and negative coupling constants $g=\pm 2\times 10^7$. In the absence of parity-odd 4PCFs, the central value of the null distribution is non-vanishing (grey curve). 
Due to the noise in the sample covariance matrix, the statistic is modified to follow a $T^2$ distribution~\citep{Sellentin2016:T2}, as discussed in \S\ref{sec:significance}.
Using either the analytic covariance matrix or the mock covariance, we obtain a similar excess in $\Delta\tilde{\chi}^2_{\rm bao}$ or $\Delta\tilde{T}^2_{\rm bao}$ relative to the null distribution.
\begin{figure*}
    \centering
    \includegraphics[width=0.8\linewidth]{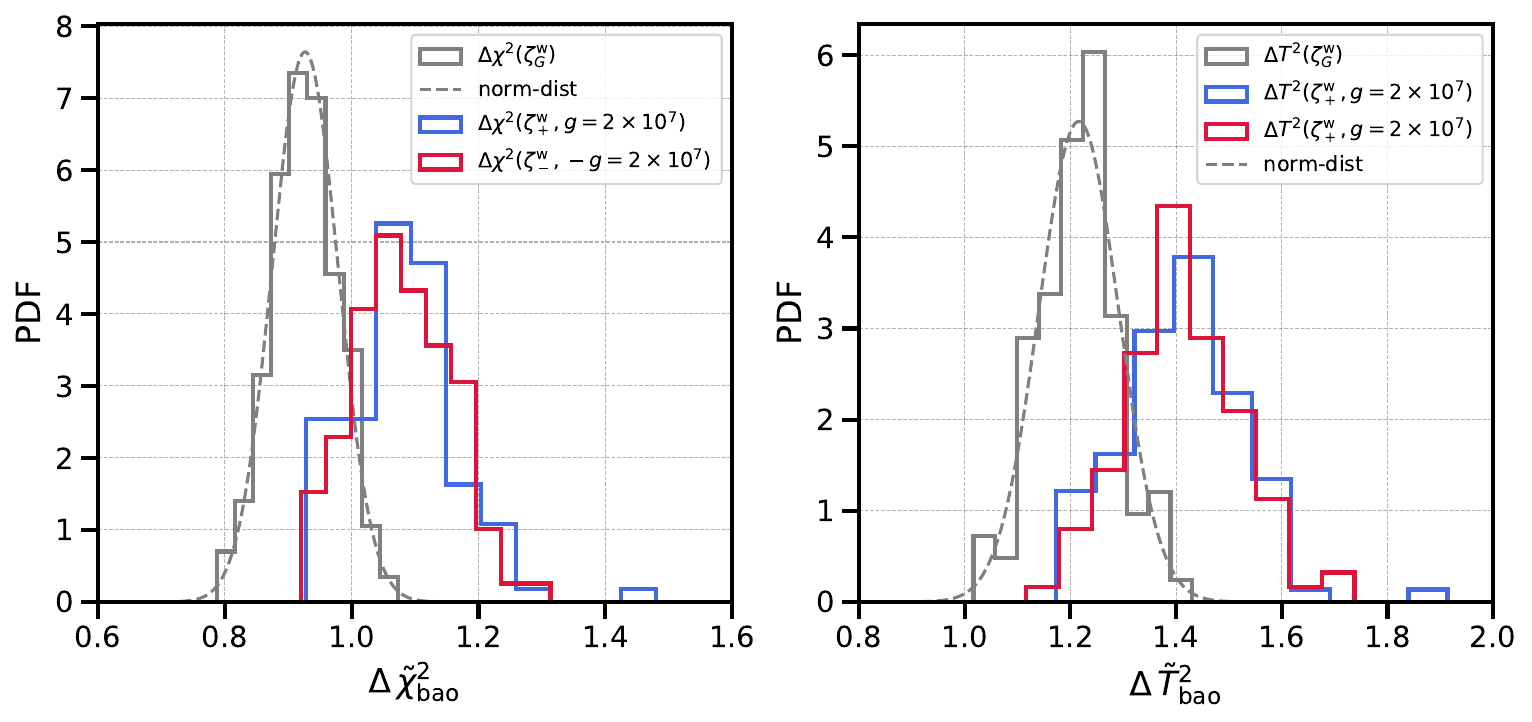}
    \caption{Distribution of the $\Delta\tilde{\chi}^2_{\rm bao}$, with analytic covariance matrix on the left panel and the mock covariance on the right panel for both coupling constant $g=\pm 2\times 10^7$. In the absence of parity-odd 4PCFs, the central value of the null distribution is non-vanishing (grey curve). Using either the analytic covariance matrix or the mock covariance, we obtain a similar excess in $\Delta\tilde{\chi}^2_{\rm bao}$ or $\Delta\tilde{T}^2_{\rm bao}$ relative to the null distribution.}
    \label{fig:stat_dist_4pcf_bao_20x160x10}
\end{figure*}

Fig.~\ref{fig:S_odd_detxn_4pcf_odd_20x160x10} and Fig.~\ref{fig:S_bao_4pcf_odd_20x160x10} show the impact of covariance on the detection significance for the overall parity-odd significance and the imprints of the BAO on the odd 4PCFs. The choice of covariance matrix does not impact the detection significance for either case. We also find the significance to be consistent for both positive and negative coupling constants.

\begin{figure*}
    \centering
    \includegraphics[width=.75\textwidth]{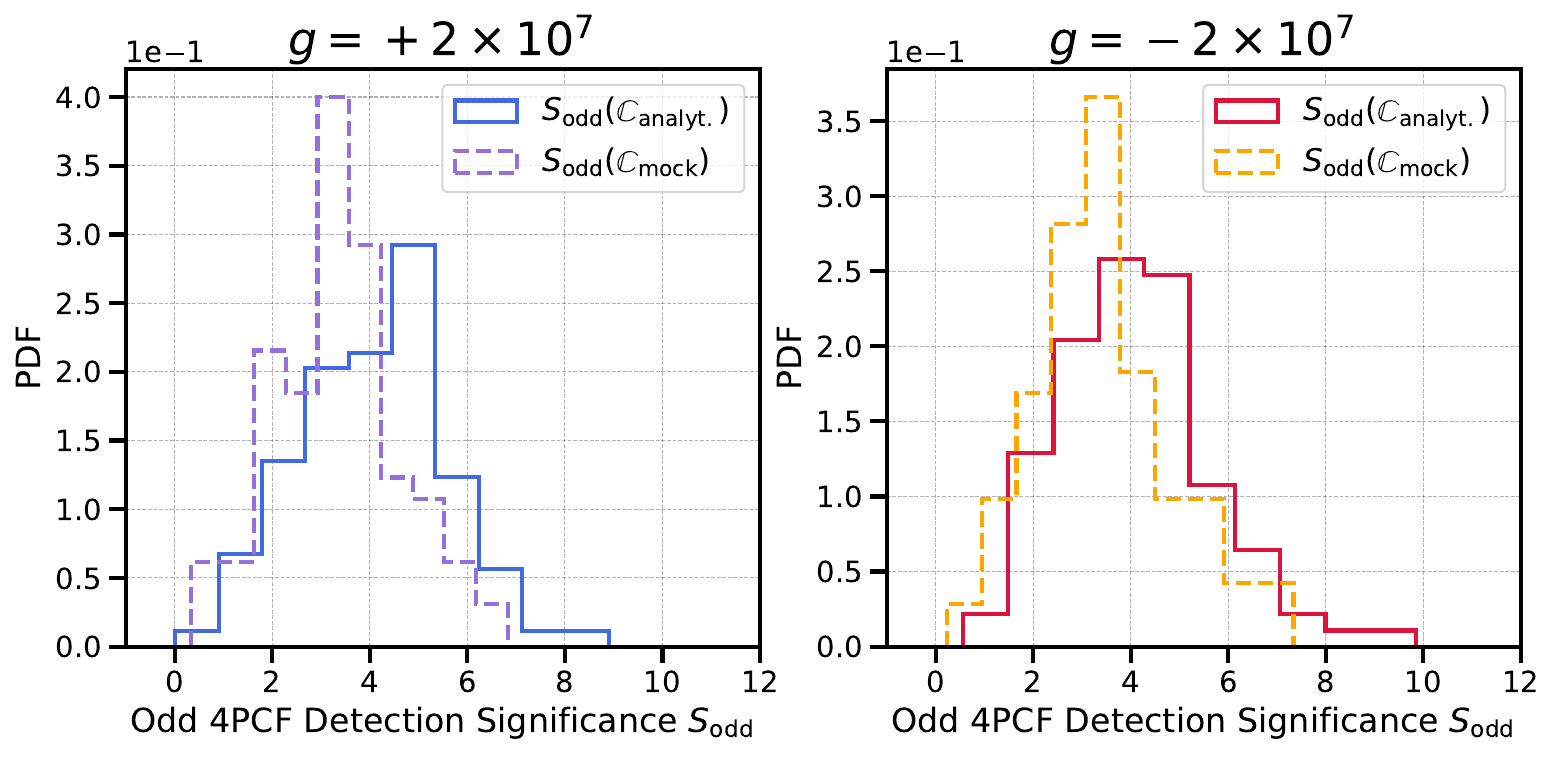}        
    \caption{Impact of the covariance matrix on the detection significance for the overall parity-odd 4PCF. The coupling constants $g=\pm 2\times 10^7$ are shown on the left and right panels, respectively. We find that the choice of covariance matrix has a marginal impact on the significance quantification. Additionally, the overall parity-odd significance is consistent for both positive and negative coupling constants.}
    \label{fig:S_odd_detxn_4pcf_odd_20x160x10}
\end{figure*}

\begin{figure*}
    \centering
    \includegraphics[width=.8\textwidth]{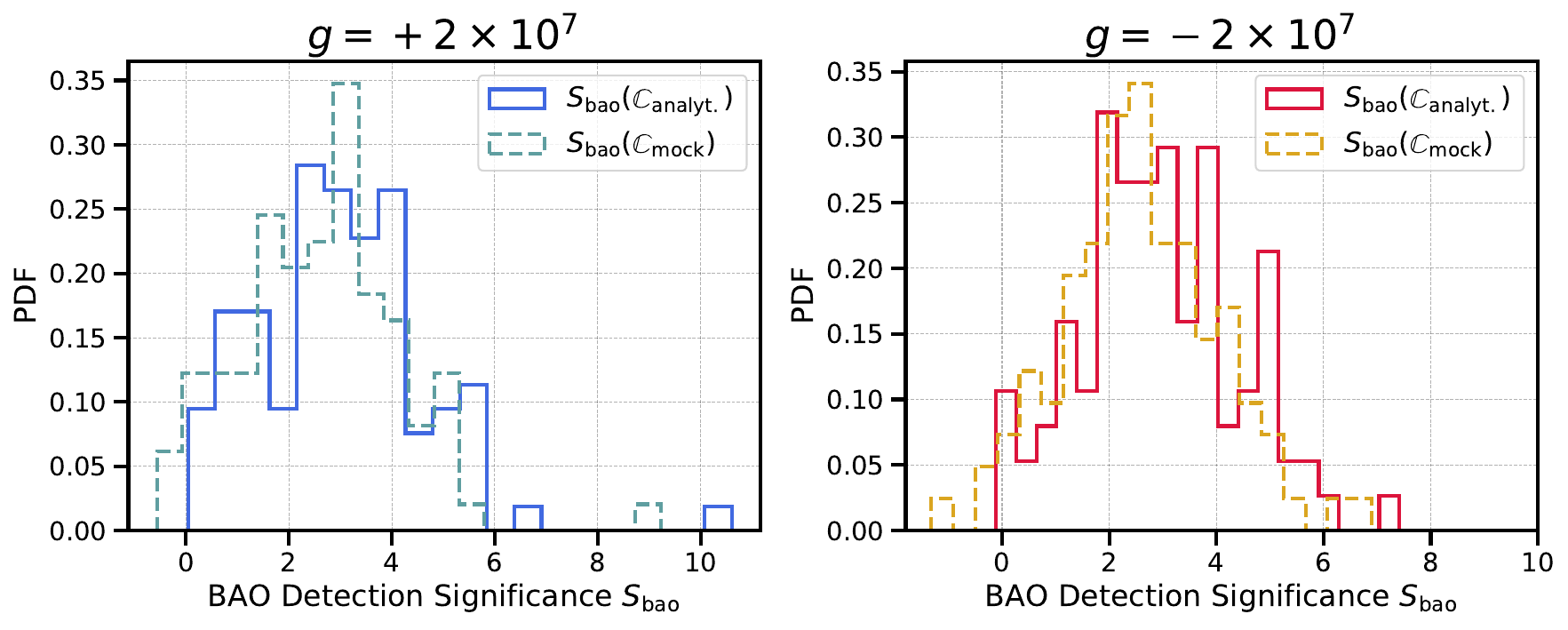}        
    \caption{Similar to Fig.\ref{fig:S_odd_detxn_4pcf_odd_20x160x10}, this plot shows the impact of the covariance matrix on the detection significance for the BAO signal on the parity-odd 4PCF. The coupling constants $g=\pm 2\times 10^7$ are shown on the left and right panels, respectively. We find that the results are independent of the choice of the covariance matrix and consistent for both positive and negative coupling constants.}
    \label{fig:S_bao_4pcf_odd_20x160x10}
\end{figure*}

% \section{Numeric Evaluation of the BAO Odd Precision}
% In this appendix, we will discuss the numbers for the BAO odd precision based on the conclusion drawn from \S~\ref{sec:discussion}.

\bibliographystyle{aipnum4-2}
\bibliography{ref.bib} 

\end{document}